\documentclass[fleqn,10pt]{wlscirep}
\title{Dynamics of Glass Forming Liquids with Randomly Pinned Particles}

\author[1]{Saurish Chakrabarty}
\author[2,*]{Smarajit Karmakar}
\author[1,3]{Chandan Dasgupta}
\affil[1]{Centre for Condensed Matter Theory, Department of Physics,
  Indian Institute of Science, Bangalore, 560012, India}
\affil[2]{TIFR Center for Interdisciplinary Science, Narsingi,
  Hyderabad 500075, India}
\affil[3]{Jawaharlal Nehru Centre for Advanced Scientific Research,
  Bangalore 560064, India.}
\affil[*]{smarajit@tifrh.res.in}


\begin{abstract}
  It is frequently assumed that in the limit of vanishing cooling
  rate, the glass transition phenomenon becomes a 
  thermodynamic transition at a temperature $T_{K}$. However, with any
  finite cooling rate, the  
  system falls out of equilibrium at temperatures near $T_g(>T_{K})$,
  implying that the very existence of the putative thermodynamic 
  phase transition at $T_{K}$ can be questioned. Recent studies of systems with  
  randomly pinned particles have hinted that the thermodynamic glass transition 
  may be observed in simulations and experiments carried out for
  liquids with randomly pinned particles. This expectation 
  is based on the results of approximate calculations that suggest that
  the temperature of the thermodynamic glass transition increases as the 
  concentration of pinned particles is increased and it may be possible to
  equilibrate the system at  
  temperatures near the increased transition temperature. We test the validity
  of this prediction through extensive molecular dynamics
  simulations of two model glass-forming liquids in the presence of random 
  pinning. We fit the temperature-dependence of 
  the structural relaxation time to the Vogel-Fulcher-Tammann form that predicts
  a divergence of the relaxation time at a temperature $T_{VFT}$ and identify
  this temperature with the thermodynamic transition temperature $T_K$.  
  We find that $T_{VFT}$ does not show any 
  sign of increasing with increasing concentration of pinned
  particles. The main effect of pinning is 
  found to be a rapid decrease in the kinetic fragility of the
  system with increasing pin concentration.  
  Implications of these observations for current theories of the
  glass transition are discussed. 
\end{abstract}
\begin{document}

\flushbottom
\maketitle
\thispagestyle{empty}

\section*{Introduction}
The glass transition is characterized by a rapid increase of the viscosity ($\eta$) and the structural relaxation time ($\tau_{\alpha}$) with decreasing 
temperature \cite{09Cav, 11BB, RFOT, RFOT1}. Recent progress 
\cite{RFOT2,10CG,14KDS,00Ediger,05Berthier,06BBMR,08BBCGV,09KDS,12KLP,12HMR,12KP,13BKP} in understanding various
dynamical aspects of this phenomenon has shed some light on this subject, but
the question of whether an ``ideal'' thermodynamic glass transition, signaled by the vanishing of the configurational
entropy density, can occur at a temperature lower than the experimentally defined (dynamic) glass transition temperature remains unanswered.
Recently, it was proposed in Ref. \cite{cammarotaPinning}, 
that the difficulty in observing the putative ideal glass transition in simulations and experiments
can be bypassed by considering liquids in the presence of quenched disorder and studying the effects
of varying disorder strength on the thermodynamic and dynamic properties of the liquid. It was
argued from Mean Field (MF) and 
Renormalization Group (RG) calculations, that a thermodynamic glass transition at a temperature higher than the transition
temperature of the liquid without disorder can
be achieved by increasing the strength of the quenched 
disorder in the system. These studies assumed that the 
Random First Order Transition (RFOT) theory \cite{RFOT, RFOT1, RFOT2},
which is currently 
the most popular theoretical framework for describing equilibrium and dynamic 
properties of glass-forming liquid near the glass transition,
remains valid in the presence of quenched disorder. 
However, the validity of the RFOT description is still controversial. 
An alternative description~\cite{kcm} of glassy dynamics, based on 
the behavior of kinetically constrained systems, does not show~\cite{jack} the occurrence 
of a transition at a non-zero temperature with increasing
disorder strength. Therefore, a numerical investigation of whether this transition actually occurs in 
model glass-forming liquids would help in determining which of these competing theories is
better at describing structural glasses. The possibility of observing a thermodynamic glass transition in 
simulations and experiments relies on the theoretically predicted increase of the transition temperature 
with increasing strength of pinning. It would be interesting to check by simulations whether this prediction
is valid for three-dimensional glass-forming liquids. A numerical study of glassy behavior in pinned liquids
would also help in understanding the results of a recent experiment~\cite{rajeshAjay2014} on colloidal 
systems with the kind of pinning disorder considered in the theoretical studies. 

Existing simulation results for the  effects of quenched disorder on the 
dynamics of supercooled liquids \cite{12KLP,13KP, 03Kim, 12BK, 13KB,14KC} 
were obtained using different ways of generating the disorder. In this study, we consider 
the disorder generated by randomly choosing 
a fraction $\rho_{pin}$ of the particles from an equilibrium configuration
of the supercooled liquid at temperature $T$ and freezing them in space. 
We henceforth refer to this geometry as the ``random pinning'' geometry.    
This kind of quenched disorder can be realized
in liquids confined in a statistically homogeneous porous medium 
\cite{krakoviackPinning} obtained by freezing a fraction of the particles
in an equilibrium configuration of the same liquid. It can also be realized in
experiments on colloidal systems~\cite{rajeshAjay2014}, using optical 
traps to pin the particles. 
This way of generating the disorder, considered in Ref.~\cite{cammarotaPinning}, has certain 
advantages. For instance, 
preparing an equilibrated system is straightforward, 
as the state obtained by instantaneously freezing a randomly selected
fraction of the particles in an equilibrated liquid configuration is a valid 
equilibrium configuration of the system with the pinning disorder. 
In earlier studies \cite{03Kim, 12KLP, 13KB,14KC}, the relaxation time of the system with
random pinning was found to increase very rapidly with
increasing concentration of the pinned particles. The RFOT description which forms the basis of the theoretical
calculations of Ref.~\cite{cammarotaPinning} predicts a divergence of the structural relaxation time at the 
putative entropy-vanishing thermodynamic transition. Therefore, the phase diagram in the
$(\rho_{pin}-T)$ plane can be obtained by locating the temperatures at which the relaxation time diverges
for different values of the pin density. Since a rapid growth of the relaxation time is a defining feature of 
glassy behavior, the dynamics of pinned liquids is interesting by itself. For these reasons, we have carried out
a detailed study of the dynamics for two model glass forming liquids in the presence of random pinning using
extensive numerical simulations.   
\begin{figure}
  \begin{center}
    \vspace{-.1in}
    \includegraphics[height=0.35\columnwidth]{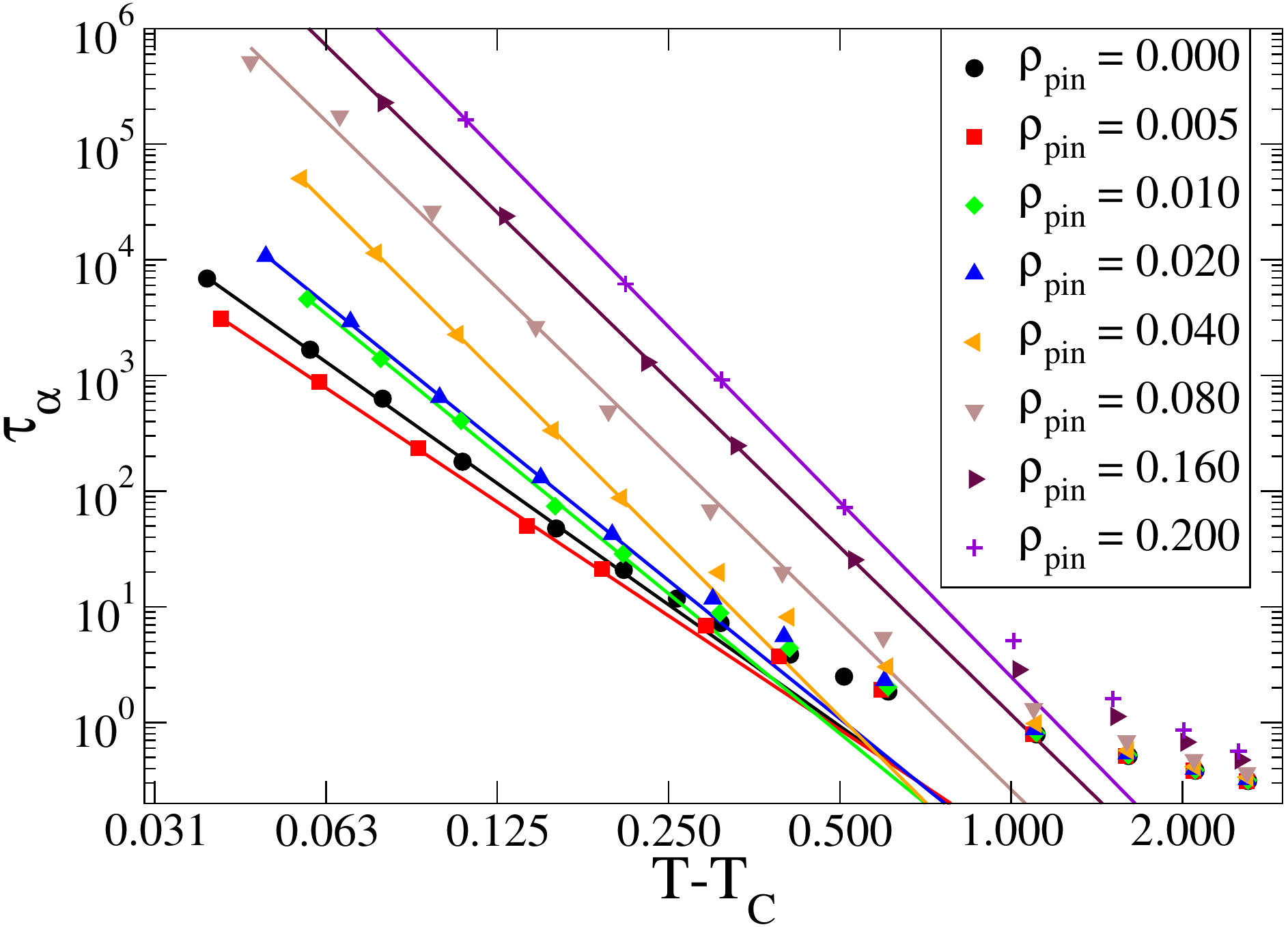}
    \hskip +1.0cm
    \includegraphics[height=0.35\columnwidth]{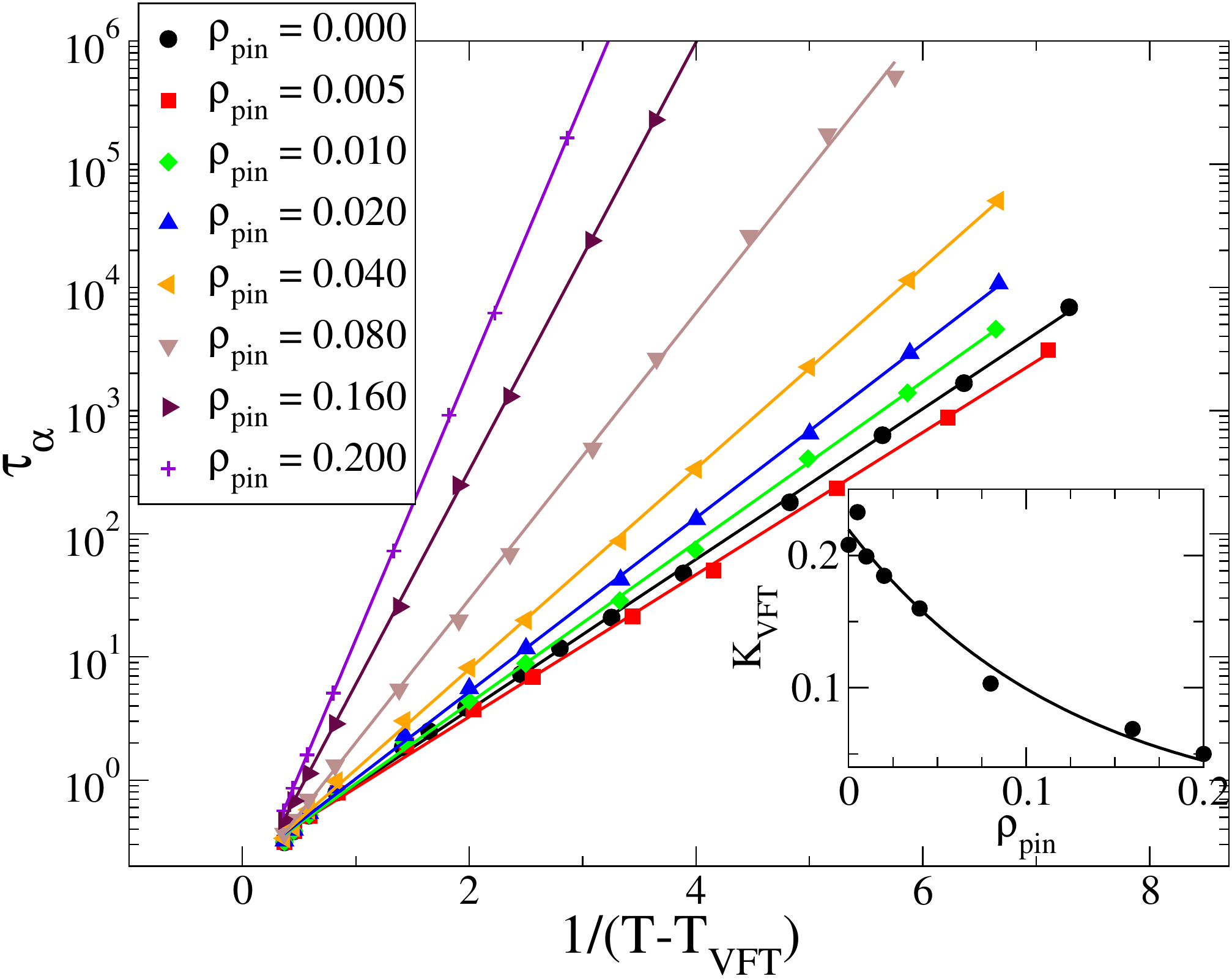}
    \caption{{\em Left Panel:} Power-law fits to obtain $T_C$ as a function
      of $\rho_{pin}$ for the 3dKA model.
      {\em Right Panel:} VFT fits to obtain $T_{VFT}$ as a function of $\rho_{pin}$ for the 3dKA model. 
      {\em Inset:} Kinetic fragility $K_{VFT}$ as a function of $\rho_{pin}$. The dramatic 
      decrease in $K_{VFT}$ with increasing $\rho_{pin}$ can be clearly seen. The line is an 
      exponential fit to the data.
      \label{mctVft3dKA}}
  \end{center}
\end{figure}
\section*{Theoretical predictions}
Before going into the details of our results, we briefly 
discuss the arguments presented in Ref. \cite{cammarotaPinning} 
about the phase diagram of the randomly pinned system
in the $(\rho_{pin}-T)$  
plane. 
In the RFOT theory, a thermodynamic glass 
transition is characterized by the vanishing of the configurational entropy 
density $s_c$ associated with the multiplicity of amorphous local minima of 
the free energy. In Ref. \cite{cammarotaPinning}, the arguments of RFOT were 
extended to glass forming system with randomly pinned particles. It is 
physically reasonable to assume that the configurational entropy of a liquid 
decreases with increase pinning fraction $\rho_{pin}$. In Ref. 
\cite{cammarotaPinning}, it was assumed that the configurational entropy 
density decreases linearly with $\rho_{pin}$,
\begin{equation}
  s_c(T,\rho_{pin}) \simeq s_c(T,0) - \rho_{pin}E(T), 
  \label{sc1}
\end{equation}   
for small values of $\rho_{pin}$, with $E(T) > 0$.  
In RFOT, $s_c(T,0)$ is supposed to go to zero at the Kauzmann
temperature $T_K(0)$, the argument of $T_K$ being the value of $\rho_{pin}$. 
Eq.( \ref{sc1}) then predicts that 
for $T>T_K(0)$, the configurational entropy should vanish at a critical pinning fraction $\rho_K(T) \simeq
s_c(T,0)/E(T)$ and a thermodynamic glass transition should occur at that pinning fraction. Assuming that this critical 
fraction $\rho_K(T) < 1$, the thermodynamic glass transition temperature $T_{K}(\rho_{pin})$, obtained from the condition 
$s_c(T_{K},\rho_{pin}) = 0$, should increase from $T_{K}(0)$ as $\rho_{pin}$ is increased from zero. 
In Ref. \cite{cammarotaPinning}, a MF analysis
was carried out for the spherical $p$-spin model 
to calculate $T_{K}(\rho_{pin})$, as well as the critical temperature ($T_C$) 
of Mode Coupling Theory (MCT)~\cite{mct} which represents the
temperature below which the  dynamics  
of the system is dominated by activated relaxation processes. It was
shown that the dependence of these two temperatures, $T_{K}$ and $T_C$,
on $\rho_{pin}$ are such that they meet at a ``critical'' value of
$\rho_{pin}$, 
at which the thermodynamic glass transition disappears. A real-space RG calculation predicted that 
the line of thermodynamic glass transitions in the $(\rho_{pin}-T)$ plane has a positive slope and it ends at a critical value of $\rho_{pin}$.
The dynamic transition at $T_C$ is not found in the RG calculation.

\begin{figure}
  \begin{center}
    \vspace{-.1in}
    \includegraphics[height=0.35\columnwidth]{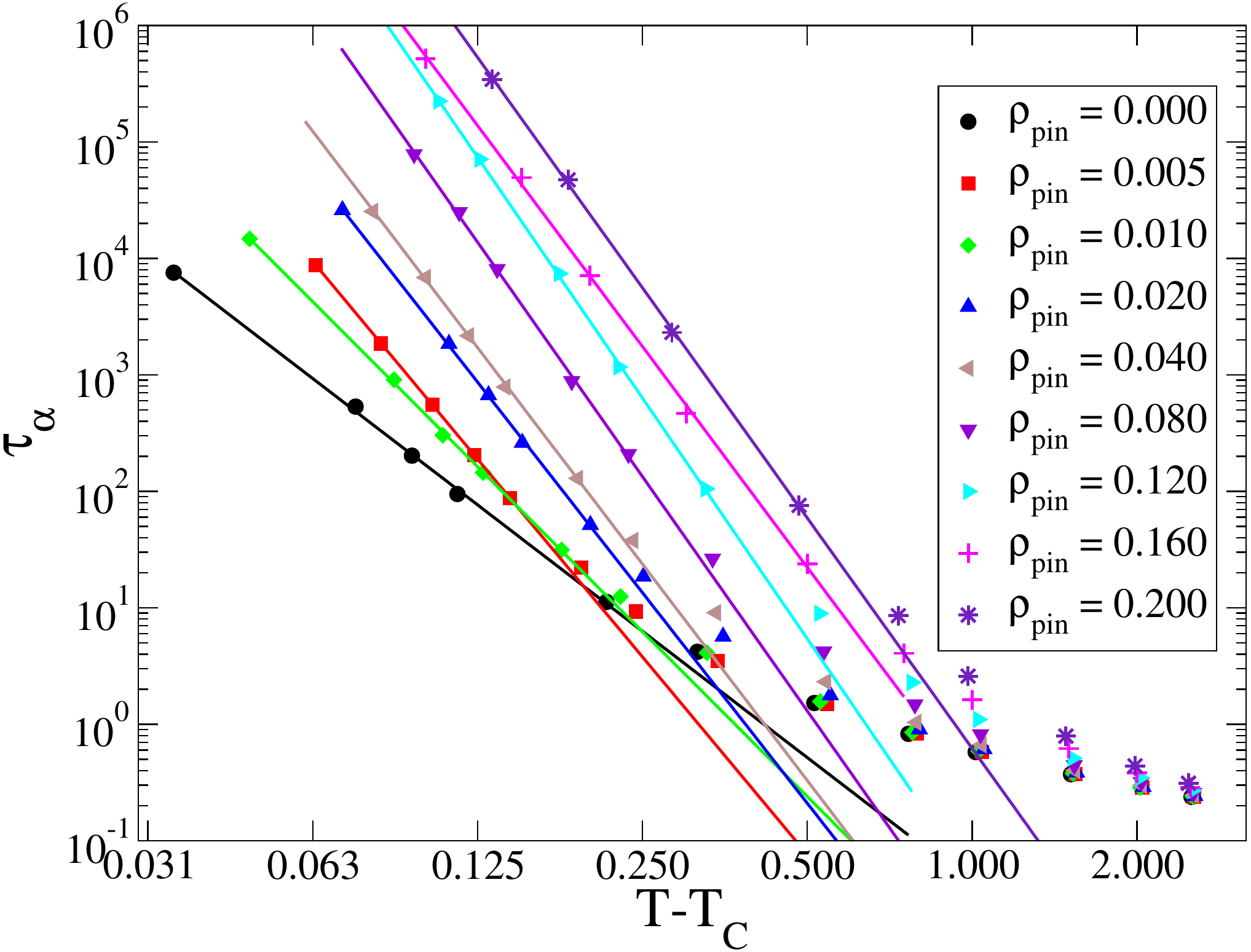}
    \hskip +1.0cm
    \includegraphics[height=0.35\columnwidth]{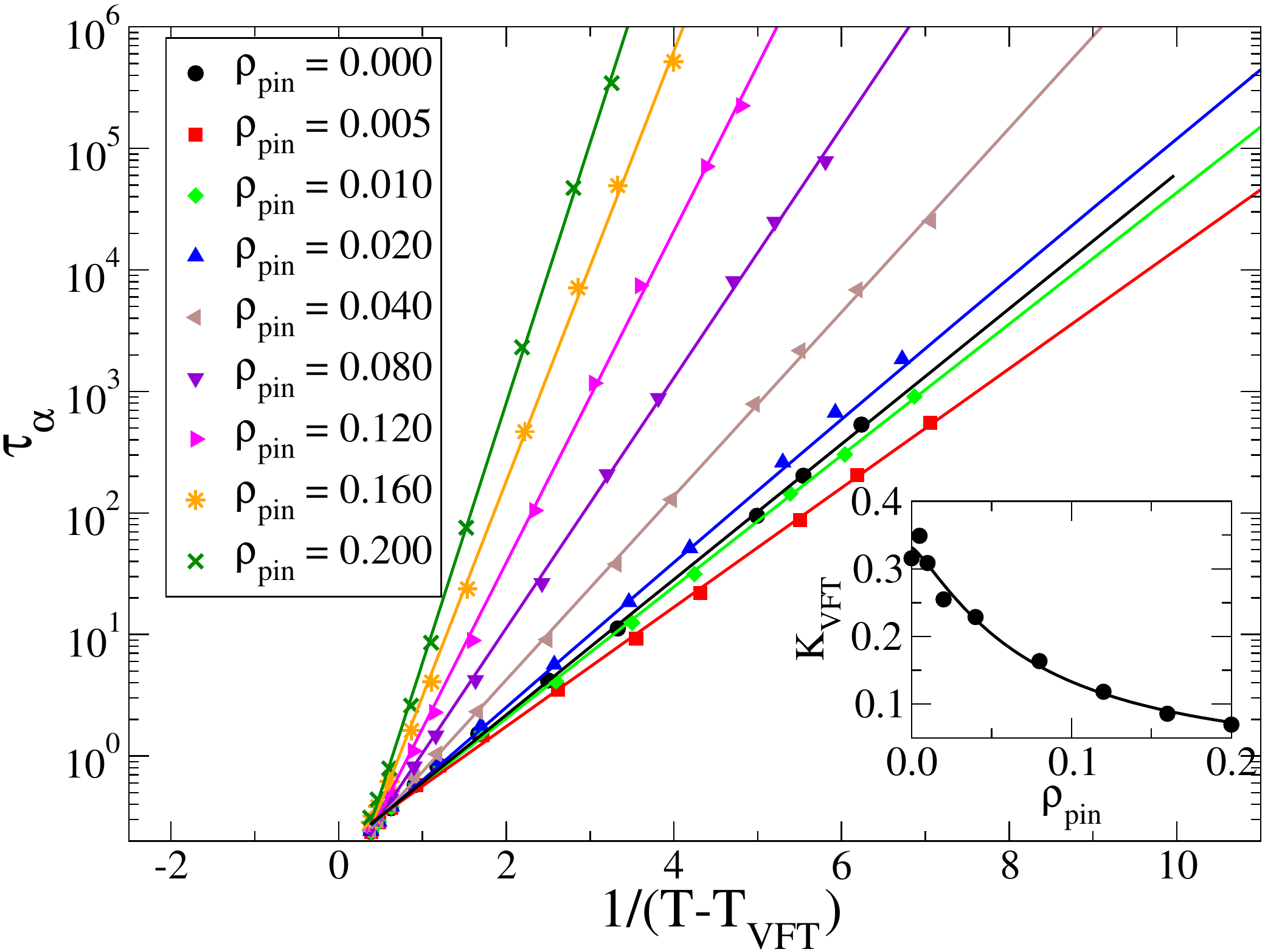}
    \caption{{\em Left Panel:} Power-law fits to obtain $T_C$ as a function
      of $\rho_{pin}$ for the 3dR10 model with $\gamma=4$.
      {\em Right Panel:} VFT fits to obtain $T_{VFT}$ as a function of $\rho_{pin}$ for the 3dR10 model. 
      {\em Inset:} Kinetic fragility $K_{VFT}$ as a function of $\rho_{pin}$. The line corresponds 
      to an exponential fit to the data.
      \label{mctVft3dR10}}
  \end{center}
\end{figure}

\section*{Systems and Methods}
The  
first model glass former we study
is the well-known Kob-Andersen \cite{95KA} $80:20$ binary Lenard-Jones 
mixture. Here it will be referred to as the {\bf 3dKA model}.  
The temperature range 
studied for this model is $[0.45, 3.00]$ at number density
$\rho = 1.20$. 
The second model studied is a $50:50$ binary mixture with pairwise 
interactions between particles that fall off with distance as an
inverse power-law  with exponent $10$ (the {\bf 3dR10 model}). 
The temperature range covered for this model is $[0.52,3.00]$ at 
number density $\rho = 0.81$. 

We performed NVT molecular dynamics
simulations using modified leap-frog algorithm with the Berendsen 
thermostat. 
For both the model systems, we performed simulations
for $\rho_{pin}$ in the range $[0.000, 0.200]$ for
each temperature. For very low temperatures, we were not able to
equilibrate the system for high pin concentrations because of a
dramatic increase in the relaxation time in these cases. (See SI for further 
details). 

\begin{figure*}
  \begin{minipage}{0.540\columnwidth}
    \hspace{-.03\columnwidth}
    \includegraphics[width=0.96\columnwidth]{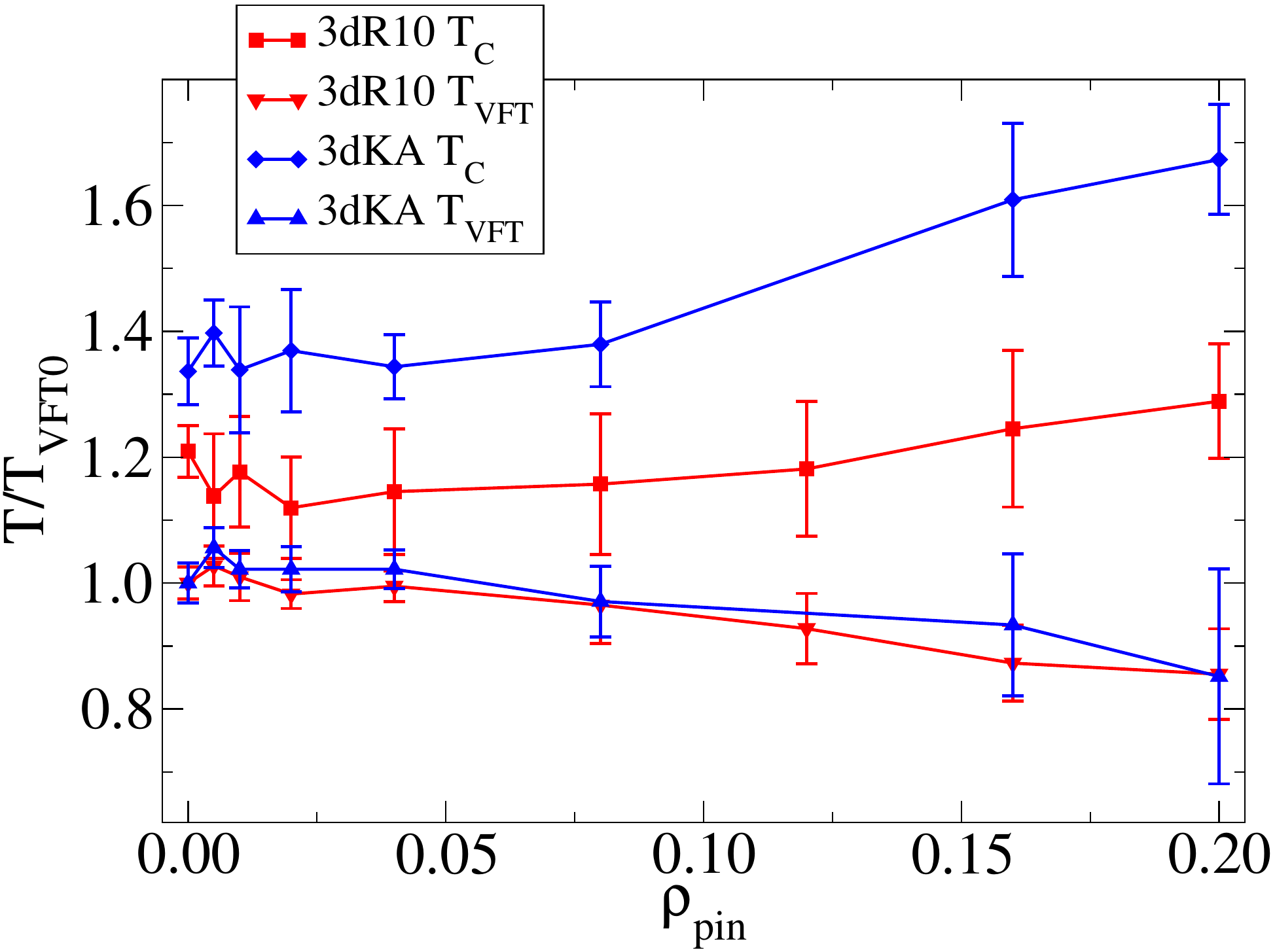}
    \vskip +0.2cm
    \caption{{\em Left panel:} Phase diagram -- Variation of the 
    MCT-transition temperature $T_C$ and the VFT-divergence temperature $T_{VFT}$
    with $\rho_{pin}$ for the 3dKA and the 3dR10 models. For
    comparison all temperatures are scaled by the VFT-divergence
    temperature for the unpinned system, $T_{VFT0}$. The lines join
    successive data points. 
    {\em Top right panel:} Angell plots for the 3dKA model. 
    {\em Bottom right panel:} Angell plots for the 3dR10 model. 
    \label{kauzMctPhaseDiagFragilityCombo}}
  \end{minipage}
  \begin{minipage}{0.450\columnwidth}
    {\includegraphics[width = 0.92\columnwidth]{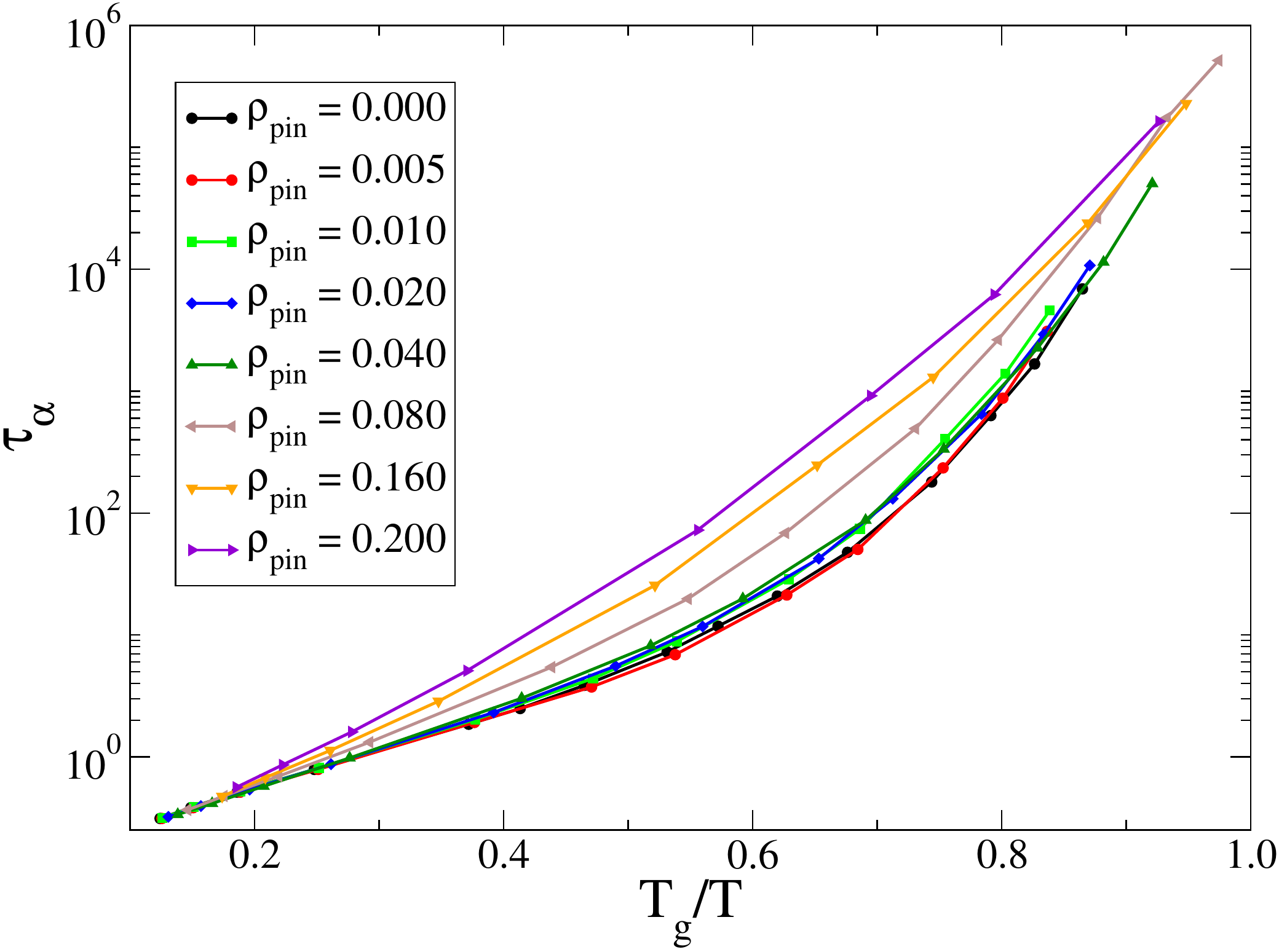}\label{fragility3dKA}}
    \vspace{.01\columnwidth}
    {\includegraphics[width = 0.92\columnwidth]{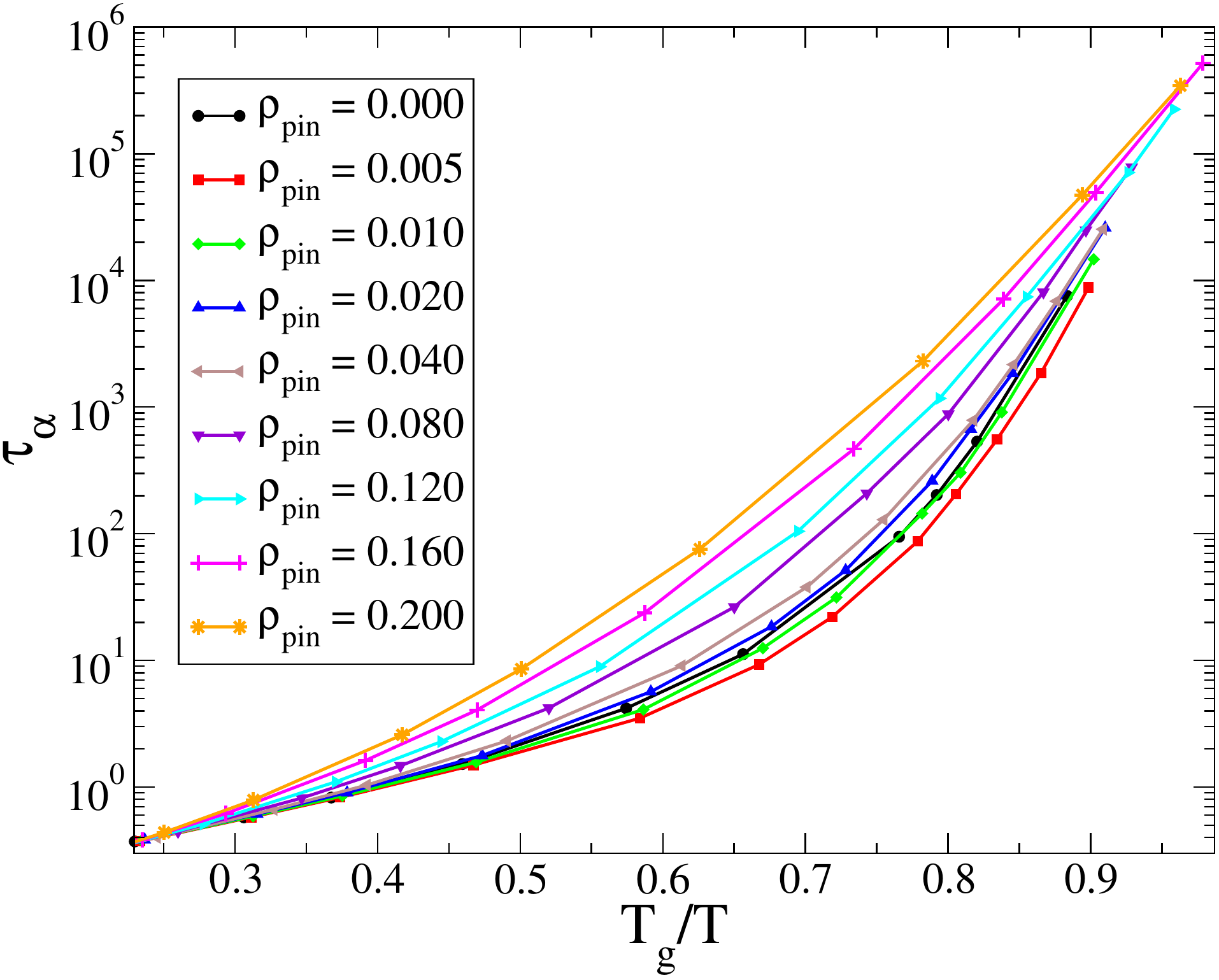}\label{fragility3dR10}}
  \end{minipage}
  
\end{figure*}
Dynamic properties are characterized by calculating the self part of a
modified two-point density correlation function which we call the overlap correlation function 
$Q(t)$ defined as
\begin{equation}
  Q(t) = \overline{\left\langle
      \frac{1}{N-N_{pin}}\sum^{'}_{i}w(|\vec{r}_i(t) - \vec{r}_i(0)|)
    \right\rangle_0}, 
  \label{overlap}
\end{equation} 
where the weight function $w(x) = 1.0$ if $x<0.30$ and $0$ otherwise,
$\overline{\langle\ldots\rangle_0}$ denotes averaging over the time origin and also averaging over different
realizations of the disorder, and $N_{pin} = \rho_{pin}N$ is the
number of pinned particles.  
The prime over the summation sign means that the sum is over only the unpinned particles. 
We average the data over $32$ independent runs for each  
state point
and consider systems with $N = 1000$. The $\alpha$-relaxation time
$\tau_{\alpha}$ is calculated by the condition $Q(\tau_{\alpha})
= 1/e$. It is to be noted that $\tau_{\alpha}$-values obtained from
the self intermediate scattering function $F_s(k,t)$, calculated at
the wave-vector at which the static structure factor $S(k)$ peaks,
are very close to those obtained using $Q(t)$. 

We estimate the MCT crossover temperature $T_C$ by fitting
$\tau_{\alpha}$-values for  
different temperatures for a given value of $\rho_{pin}$ to a
power-law form, 
$\tau_{\alpha} \sim B/|T-T_C|^{\gamma}$ and the temperature $T_{VFT}$ by
fitting the data to a Vogel-Fulcher-Tammann law defined by 
$\tau_{\alpha} \sim \tau_\infty \exp{\left[A/(T - T_{VFT})\right]}$.
It should be noted that earlier experimental and simulation studies
have reported that $T_{VFT}\approx T_K$ for unpinned liquids.
In the case of the power-law fit, for each value of $\rho_{pin}$,
4-5 data points with the highest values of 
$\tau_\alpha$ were used for the fitting procedure. The kinetic
fragility is obtained as $K_{VFT} = T_{VFT}/A$. 

\section*{Results}
Figures \ref{mctVft3dKA} and \ref{mctVft3dR10} show the 
MCT and VFT fits 
for the 3dKA and 3dR10
models respectively. One can clearly
see that the VFT fits to the data for both the model systems are excellent, but 
the MCT fits are not very good over the whole temperature range. If a few data points at relatively high
temperatures are excluded in the power-law fitting, a reasonable fit for 
low-temperature data points is obtained. This gives us confidence about the
reliability of the extracted values of $T_{VFT}$ and $T_C$, 
although both estimations rely on extrapolation. This, however, is
an unavoidable problem 
in studies of glassy dynamics, affecting both numerical and
experimental investigations.

With these caveats, if the values of $T_C$ and $T_{VFT}$ extracted from
the fits are used to construct a  
phase diagram in the $(\rho_{pin}-T)$ plane, one finds somewhat puzzling
results as depicted in the left panel of 
Fig. \ref{kauzMctPhaseDiagFragilityCombo}. This figure shows the
variation of $T_{VFT}$ and $T_C$, scaled by the value of $T_{VFT}$  
for $\rho_{pin}=0$, with $\rho_{pin}$. It is clear from the plots that 
$T_C$ increases with increasing $\rho_{pin}$, as predicted in
Ref. \cite{cammarotaPinning}  
and in subsequent detailed MCT calculations
\cite{szamelPinning,krakoviackPinning}. On the other hand, 
$T_{VFT}$ does not show any indication of increasing with $\rho_{pin}$
to meet the $T_C$ line at a critical value of $\rho_{pin}$, as
predicted in 
the MF analysis of Ref. \cite{cammarotaPinning}.
Rather, it seems to remain constant or to
decrease slowly as  $\rho_{pin}$ is increased. To cross-check these
results, we have obtained the parameters of  
the VFT form from so-called ``Stickel plots'' \cite{stickelplot}. We
find that the values of $T_{VFT}$ obtained from the lowest-temperature
data points in the Stickel plots are close to those obtained from our VFT
plots and exhibit very similar dependence on $\rho_{pin}$. The
Stickel plots  do not show any indication that this behavior will change at lower
temperatures. The details of these results are provided in 
the SI.
\begin{figure}[!h]
  \begin{center}
    \hskip -1.0cm
    \includegraphics[width=0.49\columnwidth]{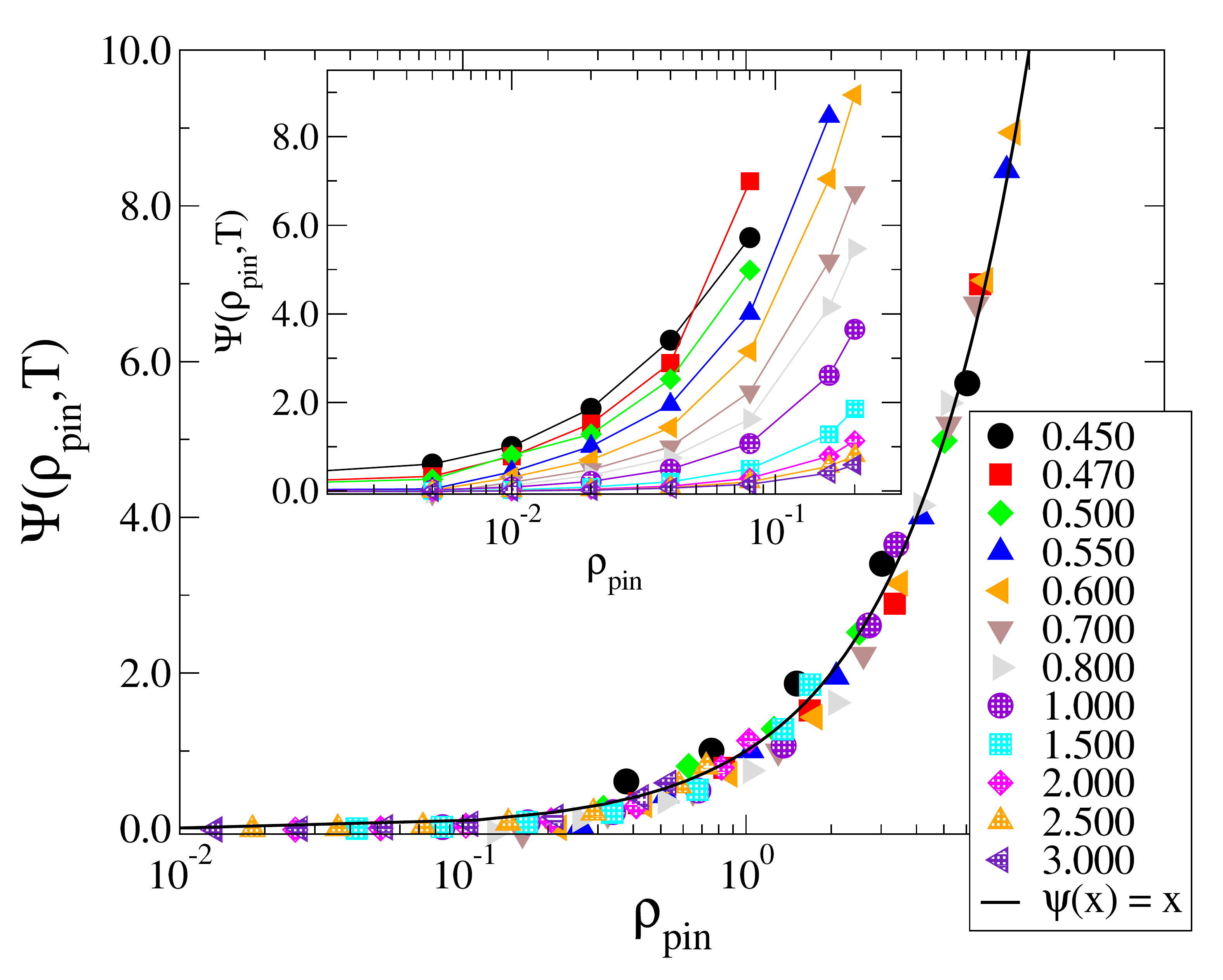}
    \hskip +0.8cm
    \includegraphics[width=0.49\columnwidth]{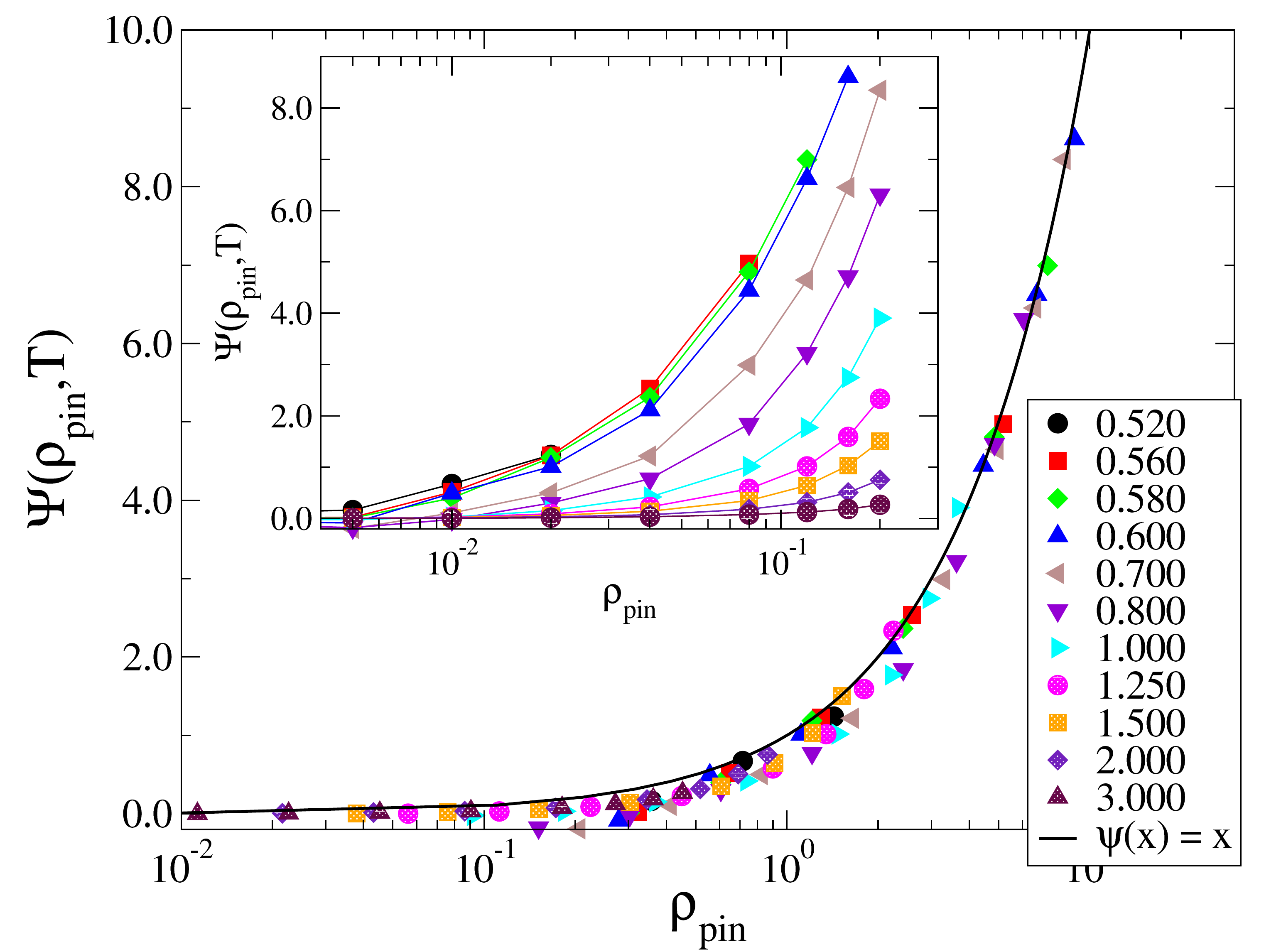}
    \caption{Data collapse using the scaling form in Eq. \ref{scalingfn} to obtain
      the length-scale $\xi_p$. Left panel is for the 3dKA system and the 
      right panel is for the 3dR10 system. In both cases, insets show 
      the uncollapsed data.
      \label{collapsexip}}
  \end{center}
\end{figure}
Another important and somewhat unexpected result of our study is that
the kinetic fragility $K_{VFT}$ decreases rapidly with increasing
$\rho_{pin}$ in both of our model systems.  
In the insets of Fig. \ref{mctVft3dKA} and Fig. \ref{mctVft3dR10},
we have plotted $K_{VFT}$ 
as a function of $\rho_{pin}$. One can see that this measure of
fragility changes by a factor of  
$5-8$ in the studied range of $\rho_{pin}$. To emphasize this point,
we have constructed ``Angell plots'' in which  
the relaxation time is plotted as a function of the temperature
scaled by $T_g$, defined as $\tau_{\alpha}(T_g) = 10^6$.  
These plots are also shown in Fig. \ref{kauzMctPhaseDiagFragilityCombo}. 
The dramatic change in
the fragility, manifested as a change in the curvature of the plots,
can be clearly seen 
in these Angell plots. Restricted fits in which $K_{VFT}$ is fixed
at the value for the unpinned liquid provide rather poor 
description of the data for relatively large values of $\rho_{pin}$
(see the SI for details).

The phenomenological Adam-Gibbs relation~\cite{adam-gibbs}, 
$\tau_{\alpha}(T) \propto \exp[B/(Ts_c(T))]$,
between the $\alpha$-relaxation time and the configurational entropy density 
($B$ is a constant) leads to a VFT form for the temperature 
dependence of $\tau_{\alpha}$ with $T_{VFT}=T_K$ if $s_c$ behaves as $Ts_c(T) =
K(T-T_K)$ where $K$ is a constant. 
Numerical results~\cite{ag-prl,sri-JCP} for the configurational entropy of liquids without pinning, 
obtained at relatively high temperatures at which the liquid can be equilibrated in time scales accessible in simulations,
are consistent with this linear relation that extrapolates to zero at a temperature $T_K$ that is close to the
$T_{VFT}$ obtained from a VFT fit to the temperature dependence of $\tau_{\alpha}$. 
If we assume that the Adam-Gibbs relation remains valid~\cite{14KC} in the presence of pinning and $s_c$ actually goes to
zero at $T=T_K=T_{VFT}$ (as in the RFOT theory), then a reduction in $T_{VFT}$ with increasing 
$\rho_{pin}$ would not be consistent with the physically reasonable 
expectation that $s_c(T,\rho_{pin})$ is a decreasing function of
$\rho_{pin}$. However, our results 
are consistent, within error bars, 
with $T_{VFT}$ being independent of $\rho_{pin}$. This can be reconciled with the requirement 
of $s_c$ decreasing with increasing $\rho_{pin}$ if the
dependence of $s_c$ on $T$ and $\rho_{pin}$ is of the form 
\begin{equation}
  Ts_c(T,\rho_{pin}) = Ts_c(T,0)F(T,\rho_{pin})=K(T-T_{VFT}) F(T,\rho_{pin}),
  \label{neweq}
\end{equation} 
where $F(T,\rho_{pin})$, the fractional reduction of the configurational entropy due to pinning,
decreases from 1 as $\rho_{pin}$ is increased from 0. If the temperature dependence of $F(T,\rho_{pin})$ is
weak (we ignore the dependence of $F$ on $T$ in the following discussion), then Eq.( \ref{neweq}) and 
the Adam-Gibbs relation would lead to a VFT form for the temperature 
dependence of $\tau_\alpha$, with $T_{VFT}$ independent of $\rho_{pin}$, in agreement with our observations. The fragility parameter
$K_{VFT}= K T_{VFT} F(\rho_{pin})/B(\rho_{pin})$ would decrease with
increasing $\rho_{pin}$ (i.e. would agree  
with our observations) if $F/B$ is a decreasing function of $\rho_{pin}$. We already know that $F$ is a decreasing function of $\rho_{pin}$. 
The requirement that $F/B$ be a decreasing function of $\rho_{pin}$  would be satisfied if $B$ increases, remains constant or decreases slower than
$F$ with increasing $\rho_{pin}$. 
In this scenario, the temperature dependence of $\tau_\alpha$ for non-zero $\rho_{pin}$ is given by
\begin{equation}
  \tau_\alpha(\rho_{pin},T) = \tau_\infty \exp\left[\frac{B(\rho_{pin})}{F(\rho_{pin})Ts_c(0,T)}\right].
\end{equation}
This implies the following relation between the relaxation times with and without pinning:
\begin{equation}
  \ln \left[\frac{\tau_\alpha(\rho_{pin},T)}{\tau_\alpha(0,T)}\right] = \frac{G(\rho_{pin})}{Ts_c(0,T)}
  \label{scaleq}
\end{equation}
where the function $G(\rho_{pin})$ is defined by
\begin{equation}
  G(\rho_{pin}) \equiv B(\rho_{pin})/F(\rho_{pin}) - B(0)/F(0).
\end{equation}
It is clear from the definition that $G(0)=0$ and assuming that $G$ and $F$ are smooth functions of $\rho_{pin}$, we get 
\begin{equation}
  G(\rho) = C\rho_{pin} + \cdots
\end{equation}
where $C$ is a constant and $\cdots$ represent terms with higher powers of $\rho_{pin}$, which can be neglected for small values of $\rho_{pin}$. So, 
for small values of $\rho_{pin}$, Eq.( \ref{scaleq}) becomes
\begin{equation}
  \ln \left[\frac{\tau_\alpha(\rho_{pin},T)}{\tau_\alpha(0,T)}\right] = \frac{C \rho_{pin}}{Ts_c(0,T)}
\end{equation}
which has the form of a scaling relation,
\begin{equation}
  \ln \left[\frac{\tau_\alpha(\rho_{pin},T)}{\tau_\alpha(0,T)}\right] = C f(\rho_{pin} \xi_p^d(T))
  \label{scalingfn}
\end{equation}
with $f(x)=x$ and the pinning length scale $\xi_p$ given by $\xi_p(T)  = [1/(Ts_c(0,T))]^{1/d} \propto [1/(T-T_{VFT})]^{1/d}$ ($d$ is the spatial dimension). 
\begin{figure}
  \begin{center}
    \vspace{-0.5cm}
    \hskip -0.80cm\includegraphics[width=0.45\columnwidth]{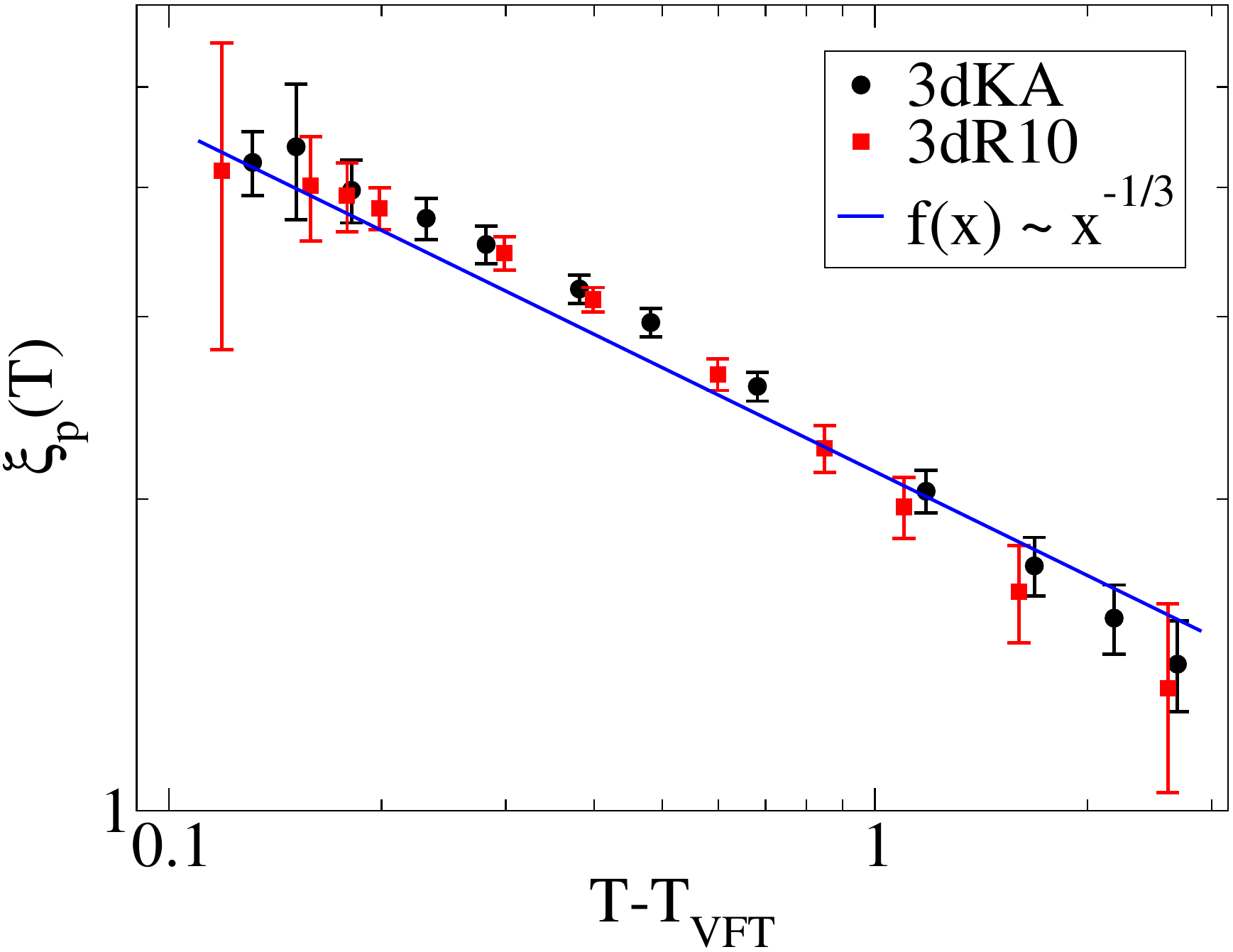} 
    \caption{Pinning length scale $\xi_p$ as a function of temperature $T$.
      \label{xiPdvsT}}
  \end{center}
\end{figure}

To test this scaling prediction, we have tried to collapse the 
data for $\psi(\rho_{pin},T) \equiv \ln[\tau_\alpha(\rho_{pin},T)/\tau_\alpha(0,T)]$ for all temperatures and pinning densities into
a single scaling curve by choosing the length scale $\xi_p(T)$ appropriately for different temperatures. As shown in 
Fig. \ref{collapsexip}, good data collapse is obtained for both the model systems studied. The line passing 
through the collapsed data is indeed of the form predicted by the 
scaling argument. 
We have also checked whether the $\xi_p(T)$ obtained from the 
scaling collapse is proportional to $ [1/(T-T_{VFT})]^{1/d}$. In Fig. \ref{xiPdvsT}, we have plotted the pinning length scale
$\xi_p$ as a function of $\left(T - T_{VFT}\right)$ for both the model
systems. The results are clearly consistent with the scaling
argument. It is interesting to note that although our results for the phase diagram do not agree
with the prediction of Ref.~\cite{cammarotaPinning}, the temperature dependence of the pinning length $\xi_p(T)$  turns out 
to be the same as that in Ref.~\cite{cammarotaPinning}. 

It is also possible that $F(\rho_{pin})$ goes to zero 
at a value of $\rho_{pin}$ higher than the largest value 
considered in our simulations. If this happens, then the line of (putative) 
thermodynamic glass transitions in the $(T-\rho_{pin})$ plane would end at 
this ``critical'' value of $\rho_{pin}$. This would be similar to the phase diagram 
obtained in the RG calculation reported in Ref~\cite{cammarotaPinning}, with 
the important difference that the transition line would be parallel to the $\rho_{pin}$ axis. 
It should, however, be noted that the arguments above do not depend on the assumption that 
a thermodynamic glass transition arising from the vanishing of $s_c$ actually occurs in liquids 
without pinning. Since numerically obtained values of $s_c(T,0)$ at the relatively high temperatures
considered in our simulations satisfy the relation $T s_c(T,0) = K(T-T_{VFT})$ irrespective of
whether $s_c$ actually goes to zero at a non-zero temperature, 
our results are also compatible with 
scenarios, such as those based on the behavior of kinetically constrained 
models~\cite{kcm}, in which 
a thermodynamic glass transition does not occur at any non-zero temperature. In such a scenario,
a slow reduction of the value of $T_{VFT}$,
obtained from fitting the simulation data 
for $\tau_\alpha$ to the VFT form, with increasing pin concentration
would not be incompatible with the physical
requirement of $s_c$ decreasing with increasing $\rho_{pin}$.

\section*{Comparison with Other Numerical Results}
Our results for the phase diagram in the $(\rho_{pin}-T)$ plane, with
$T_{VFT}$ identified with $T_K$, are in disagreement with those 
of two recent numerical studies~\cite{13KB,OKIM14}. 
In Ref. \cite{13KB}, the dependence of $T_{K}$ on $\rho_{pin}$ was
obtained from simulations of a 64-particle system of harmonic
spheres. The temperature 
$T_{K}$, obtained from the behavior of the 
distribution of an overlap parameter similar to that defined in Eq.( \ref{overlap}), was found to increase with increasing $\rho_{pin}$. 
We believe that the results reported in Ref.\cite{13KB} suffer from strong finite-size effects. 
As shown in that paper,  the basic form of the distribution (whether it is unimodal or bimodal) for a system with 
64 particles can be different from that for a system with 128 particles. Similar results for strong finite-size effects in the distribution of
a similar overlap parameter were reported earlier~\cite{09KDS}.
Since a change in the distribution from unimodal to bimodal is supposed to signal the transition from the liquid to the glass 
phase,  the data for the distribution obtained for a system with 64 particles cannot provide reliable quantitative information 
about the location of the transition point in the $(\rho_{pin}-T)$ plane. This, we believe, is the primary reason for the difference
between our results and those of Ref.\cite{13KB}.
We have also found that the fragility for a 64-particle system (3dR10 model)
is substantially smaller than that for a system with 1000 particles. Since our 
results show that the main effect of pinning is a reduction of the
fragility, studies of small systems in which the fragility is strongly affected by system size 
are not expected to provide a reliable description of the effects of pinning.  
\begin{figure*}
    \begin{center}
      \vskip -1.00cm
      \begin{minipage}{0.52\columnwidth}
      \includegraphics[height=0.74\columnwidth]{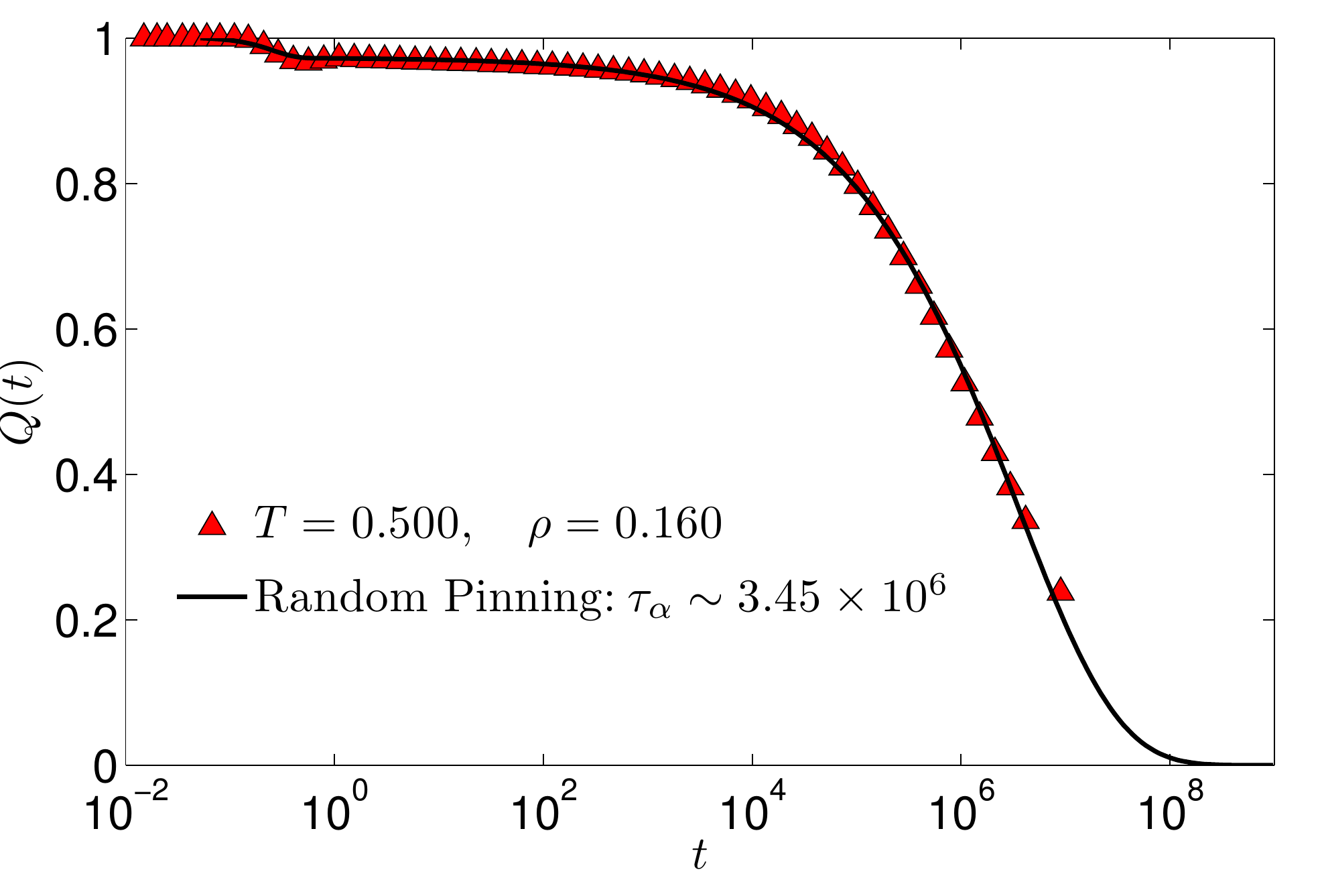}
      \caption{Overlap correlation function $Q(t)$ for three different 
        state points, ($T=0.500$, $\rho_{pin}=0.160$), 
        ($T = 0.550$, $\rho_{pin} = 0.200$) and ($T = 0.700$, $\rho_{pin}
        = 0.300$), near the phase boundary obtained in
        Ref.\cite{OKIM14} from vanishing of the  
        configurational entropy. The system size, $N = 300$, is the same as that in
        Ref.\cite{OKIM14}. Results for $Q(t)$ obtained using the template protocol 
        for selecting the pinned particles are
        shown in the middle and right panels for comparison. The lines are fit of
        the $Q(t)$ data using the functional form 
        $Q(t) = Ae^{-(t/\tau_b)^2} + Be^{-(t/\tau_{\alpha})^{\beta}}$ with 
        $A, B,\tau_b,\tau_{\alpha},\beta$ as variables.
        \label{overlapCorl}}
      \end{minipage}
      \qquad
      \begin{minipage}{0.42\columnwidth}
      \includegraphics[height=0.750\columnwidth]{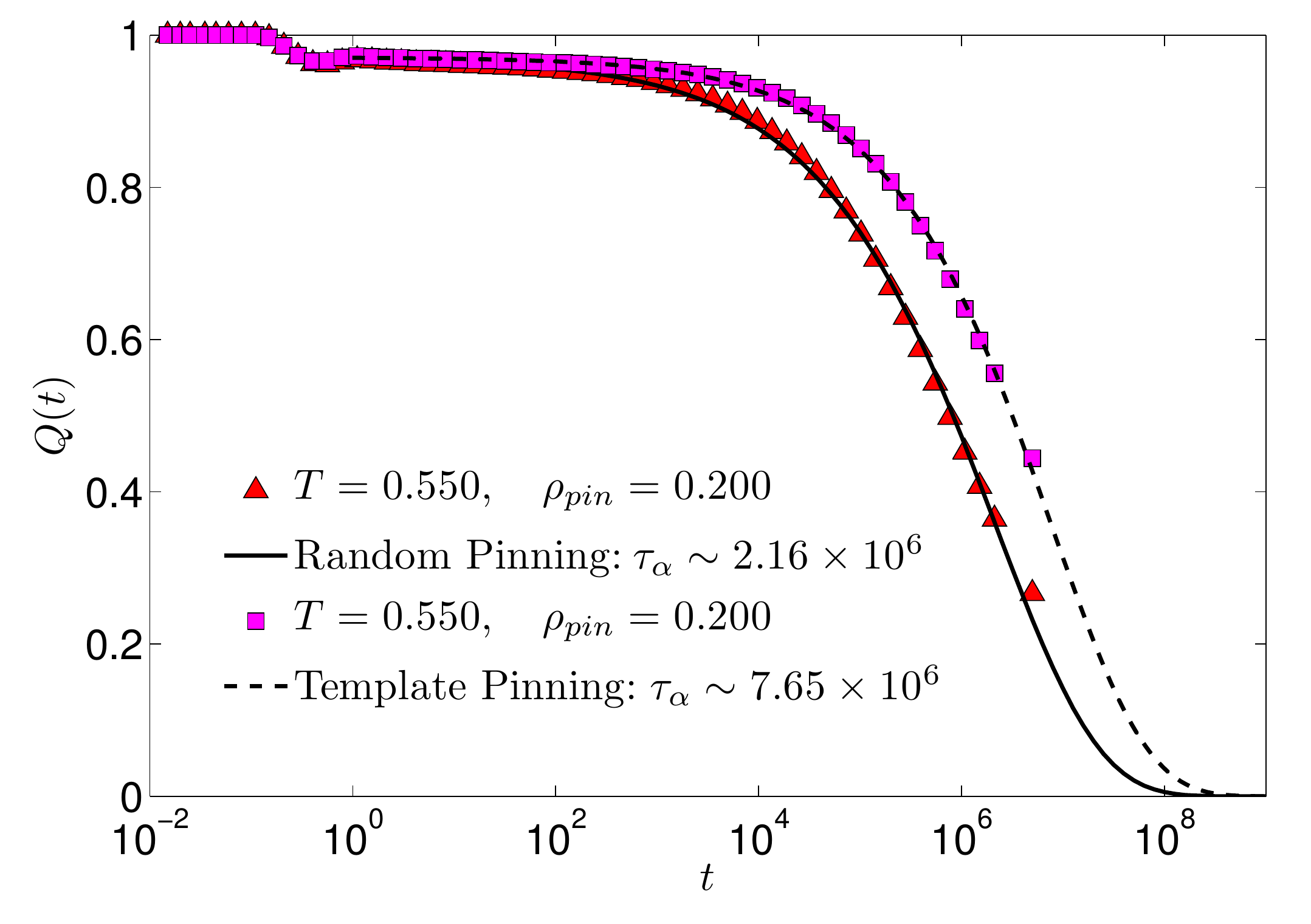}
      \includegraphics[height=0.750\columnwidth]{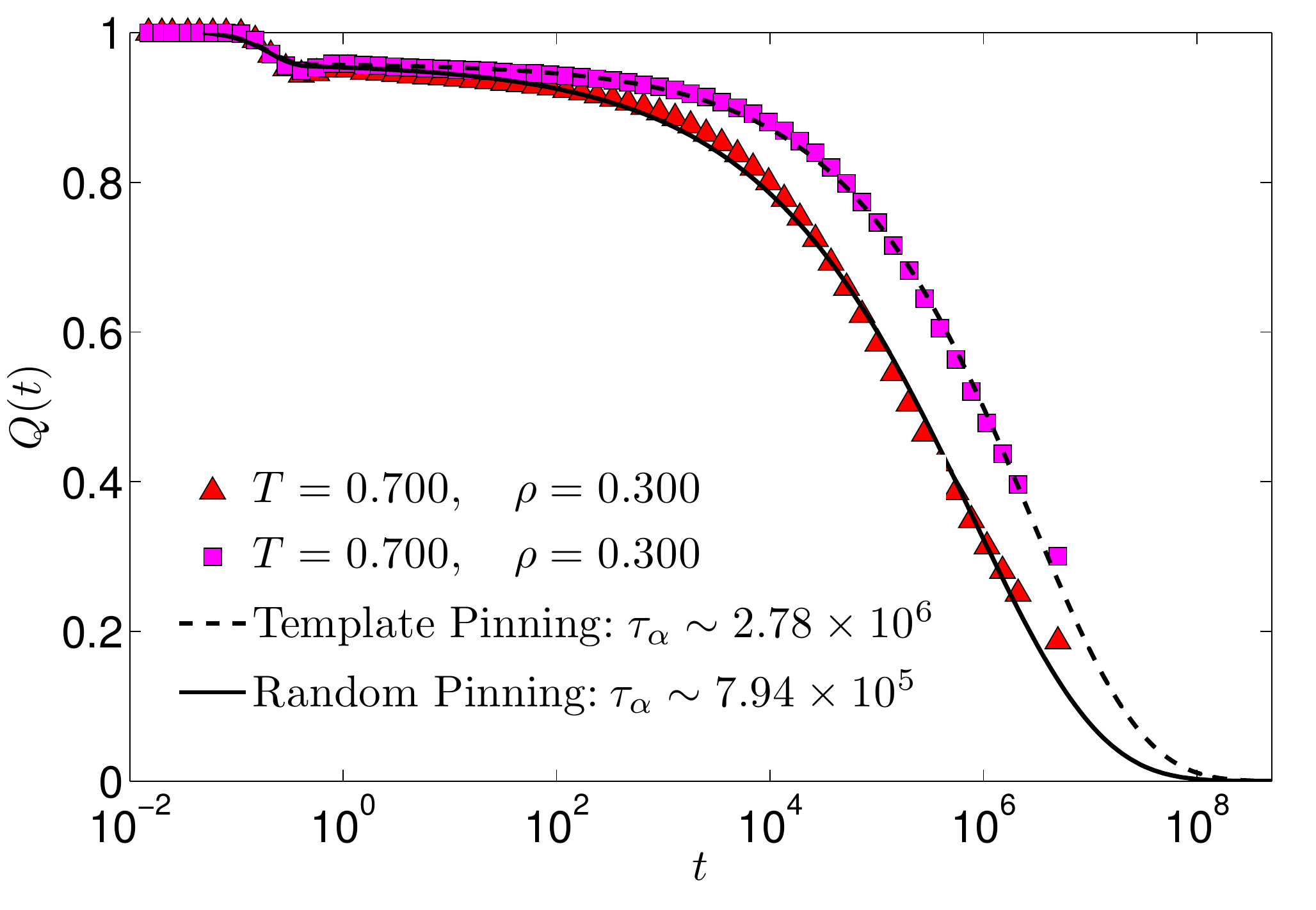}
      \end{minipage}
    \end{center}
  \end{figure*}
In another recent study~\cite{OKIM14},  the configurational entropy density $s_c$ of the 3dKA
model was calculated for different pin densities and temperatures
and the phase boundary in the $(\rho_{pin}-T)$ plane was obtained by estimating the values of $\rho_{pin}$ at which  
$s_c$ goes to zero for different temperatures. The value of $T_{K}$ obtained in this way was found to
increase with increasing pin concentration. These results, obtained from the behavior of thermodynamic quantities, are quite different from those obtained in our work from the dynamics of
the system. 

  To resolve this discrepancy,
  we have simulated the dynamics of the 3dKA model at a few points in the $(\rho_{pin}-T)$  plane at which $s_c$
  is supposed to go to zero according to the phase diagram obtained in Ref.\cite{OKIM14}.  
  In these simulations, we considered the same system size ($N = 300$) as that in 
  Ref.~\cite{OKIM14} to remove any ambiguity that may arise from finite-size 
  effects (we found similar results for $N=1000$, indicating that finite-size effects for the dynamics are weak 
  for systems with 300 or more particles).
  Surprisingly, we found that it is possible to equilibrate the system in time scales accessible in MD simulations at the points where $s_c$
  is supposed to go to zero or to have a very small value. Results for the overlap function $Q(t)$ at three ``transition points'',
  ($T=0.50$, $\rho_{pin}=0.16$), 
  ($T=0.55$, $\rho_{pin}=0.20$) and 
  ($T=0.70$, $\rho_{pin}=0.30$), are shown in  Fig. \ref{overlapCorl}.
  The relaxation times at these points are estimated to be of the order of 
  $10^6-10^7$,  as indicated in the legends of the respective plots. This is very different from the behavior of  systems without pinning, for which
  it is impossible to equilibrate the system in time scales accessible in MD simulations at temperatures close to the
  value at which $s_c$ is supposed to go to zero. For example, $\tau_\alpha$ in the unpinned 3dKA model attains the value of
  $10^7$ at a temperature that is 25\% higher than the temperature at
  which $s_c$ extrapolates to zero. 

  These results seem to imply that the relaxation time is {\it finite} at points in the $(\rho_{pin}-T)$  plane where $s_c$ goes
  to zero (or has a very small value) according to the results reported in Ref.~\cite{OKIM14}. While the
  possibility that the relaxation time does not diverge when the configurational entropy 
  vanishes cannot be ruled out, 
  we believe that this is unlikely to be the explanation of the seemingly contradictory results mentioned above.  
  The dependence of $\tau_\alpha$ on $s_c$ may not be the Adam-Gibbs 
  relation (it is suggested in Ref.~\cite{OKIM14}
  that the Adam-Gibbs relation is violated in pinned systems), but a violation of the physically reasonable expectation 
  that $\tau_\alpha$ should diverge if $s_c$ goes to zero would be quite surprising. A more probable explanation of these
  results is that $s_c$ does not actually go to zero at the phase boundary obtained in Ref.~\cite{OKIM14}.
  The calculation of $s_c$ in Ref.~\cite{OKIM14} requires estimation of the ``basin
  entropy'' which is calculated using a harmonic approximation. Anharmonic effects at the relatively high temperatures 
  considered for pinned systems may cause inaccuracies in the estimation of $s_c$, leading to errors in the determination 
  of the phase boundary in the $(\rho_{pin}-T)$  plane. One of the signatures
of this effect has been pointed out by the authors of Ref. \cite{OKIM14} as
the negative values of the configurational entropy at some state
points. Another point to note is that the
  pinned particles were chosen randomly in our simulations, whereas a ``template'' method was used in Ref.~\cite{OKIM14}. We have
  found that time scales obtained for template pinning are systematically larger than those for randomly pinned systems 
  at the same temperature and pin density. Typical results for 
  ($T=0.55$, $\rho_{pin}=0.20$) and 
  ($T=0.70$, $\rho_{pin}=0.30$) are shown in the middle 
  and right panels of  Fig. \ref{overlapCorl}. 
  It is not clear whether this difference between the protocols used for selecting the pinned particles would account for the difference between 
  our results and those of Ref.~\cite{OKIM14}. It would be interesting 
  to find out whether the configurational entropy is affected by the choice of the protocol. 
  It is argued in Ref.~\cite{OKIM14} that template pinning  reduces  sample-to-sample fluctuations compared 
  to random pinning, but we found that fluctuations in the overlap function are of similar 
  magnitude for both protocols. It is also possible (but very unlikely) that the
  plots for $Q(t)$ in Fig. \ref{overlapCorl} would level off 
  at time scales longer than those considered in our simulations, leading to very large values of the relaxation time. The excellent fits of the
  data for $Q(t)$ to a stretched exponential form, shown in Fig. \ref{overlapCorl}, strongly argues against this possibility.

In a recent numerical study~\cite{jcp2015} of the dynamics of the Kob-Andersen mixture 
in the presence of random pinning of the kind considered here, it was found, from VFT fits 
to the data for $\tau_\alpha$ as a function of $\rho_{pin}$ at a fixed temperature $T$,
that the values of $\rho_{pin}$ at which $\tau_\alpha$ appears to diverge are large ($\simeq 0.58$)
and essentially independent of $T$ for $T \leq 0.7$. These ``critical'' values of 
$\rho_{pin}$ are very different from those predicted in the phase diagram of Ref.~\cite{OKIM14}.
The dynamic behavior found in Ref.~\cite{jcp2015} is qualitatively similar to that found
in our study and consistent with the behavior of the configurational entropy proposed in 
Eq.(\ref{neweq}) with $F(\rho_{pin})$ going to zero at $\rho_{pin} \simeq 0.58$.

\section*{Conclusion}
In summary, we have obtained the phase diagram of two model glass forming 
liquids with randomly pinned particles from a study of the temperature dependence 
of the structural relaxation time and found that the MCT temperature 
$T_C$ increases, in agreement with the predictions of Ref.
\cite{cammarotaPinning}, whereas the VFT-divergence temperature $T_{VFT}$ remains nearly 
constant or decreases slowly with increasing pin concentration. The second 
observation is important because it is in disagreement with the 
predictions of 
Refs. \cite{cammarotaPinning,13KB,OKIM14}
if we assume (as is done in the RFOT description) that a thermodynamic
transition at which the configurational entropy density $s_c$ goes to zero
coincides with a divergence of the relaxation time.
If we interpret our results as 
indicating that $T_{VFT}$ decreases with increasing pin concentration, then the 
unavoidable conclusion would be that the thermodynamic glass transition of 
RFOT {\em does not} take place in these systems. An interpretation of our results as 
showing that $T_{VFT}$ is independent of the pin concentration would imply that 
{\bf (a)} the RG calculation of Ref.\cite{cammarotaPinning} is not 
quantitatively accurate and that {\bf (b)} the dependence of $T_{VFT}$
on $\rho_{pin}$ 
found in Refs. \cite {13KB} and \cite{OKIM14} is not quantitatively accurate. 
In either case, the original expectation that the 
addition of quenched random pinning would make the thermodynamic glass 
transition accessible to experiments and simulations would not be fulfilled.
It is also possible (though unlikely in our opinion) that a vanishing of $s_c$
does not correspond to a divergence of $\tau_\alpha$ in the pinned systems
considered here. This would imply that the RFOT description does not apply to 
these systems.

These findings will help in interpreting the results of 
experiments \cite{rajeshAjay2014} on colloidal systems with random pinning. 
We also find a rapid reduction of the kinetic fragility with increasing pin 
concentration. Our results indicate that a reduction of the fragility, 
rather than an increase in the VFT-divergence temperature, is responsible for the 
increase in the relaxation time with increasing pin
concentration. Since the fragility changes by factor of $5-8$ 
as the pin concentration is changed, model liquids with randomly pinned 
particles may also be useful for understanding the role of fragility in the 
glass transition.

\section*{Acknowledgments}
We would like to thank Srikanth Sastry for useful discussions. S.C.
wishes to thank the UGC's Dr. D.S. Kothari Fellowship for financial support
and TCIS for hospitality. 

\section*{Author contributions statement}
S.C. and S.K. performed simulations and analyzed data. S.K and
C.D. guided the research and provided the main ideas. All authors
wrote the manuscript.

\bibliography{$HOME/Dropbox/docs/mybiblio.bib}

\end{document}


\title{ Phase Diagram of Glass Forming Liquids with Randomly Pinned
  Particles -- Supplementary Information}

\author{Saurish Chakrabarty${^1}$}
\author{Smarajit Karmakar$^{2}$}
\author{Chandan Dasgupta$^{1,3}$}
\affiliation{$^1$ Centre for Condensed Matter Theory, Department of Physics,
Indian Institute of Science, Bangalore, 560012, India, $^2$ Centre for Interdisciplinary Sciences, Tata Institute of 
Fundamental Research, 21 Brundavan Colony, Narisingi, Hyderabad, India, $^3$ Jawaharlal Nehru Centre for Advanced 
Scientific Research, Bangalore 560064, India.}

\maketitle

\maketitle
\section{Model and Simulation}
The  first model glass former we study is the well-known Kob-Andersen \cite{95KA} $80:20$ binary 
Lenard-Jones mixture. In the main article it is referred to as the 3dKA model. The interaction 
potential in this model is given by 
\begin{equation}
V_{\alpha\beta}(r)=4\epsilon_{\alpha\beta}\left[\left(\frac{\sigma_{\alpha\beta}}{r}\right)^{12}-
\left(\frac{\sigma_{\alpha\beta}}{r}\right)^{6}\right],
\end{equation}
where $\alpha,\beta \in \{A,B\}$ and $\epsilon_{AA}=1.0$, $\epsilon_{AB}=1.5$,
$\epsilon_{BB}=0.5$, $\sigma_{AA}=1.0$, $\sigma_{AB}=0.80$,
$\sigma_{BB}=0.88$. The interaction potential is cut off at $2.50\sigma_{\alpha\beta}$ and we
use a quadratic polynomial to make the potential and its first two derivatives smooth at the
cutoff distance. The temperature range 
studied for this model is $T \in \{0.450, 3.000\}$ at number density $\rho = 1.20$ . 
The second model studied is a $50:50$ binary mixture with pairwise interactions between 
the particles that falls off with distance as an inverse power law  as
\begin{equation}
V_{\alpha\beta}(r) = \epsilon_{\alpha\beta}\left(\frac{\sigma_{\alpha\beta}}{r}\right)^n,
\end{equation}
with exponent $n = 10$ (the 3dR10 model). The potential is cut off at $1.38\sigma_{\alpha\beta}$. 
We  again use a quadratic polynomial to make the potential and its first two derivative smooth at the 
cutoff. The parameters of the potential are: $\epsilon_{\alpha\beta} = 1.0$, $\sigma_{AA} = 1.0$,
$\sigma_{AB} = 1.22$ and $\sigma_{BB} = 1.40$. The temperature range covered for this model is
$T \in \{0.52,3.00\}$ at number density $\rho = 0.81$.

NVT molecular dynamics simulations are done in a cubic simulation box with periodic
boundary conditions in three dimensions for both the model systems. We use the modified 
leap-frog algorithm with the Berendsen thermostat to keep the temperature constant in the simulation runs.  
Length, energy and time scales are measured in units of $\sigma_{AA}$, $\epsilon_{AA}$ and
$\sqrt{\sigma_{AA}^2/\epsilon_{AA}}$.  The integration time step used is $dt = 0.005$ 
in this temperature range. Equilibration runs are performed for $\sim 10^8 - 10^9$ MD steps depending 
on the temperature and production runs are long enough to ensure that the two-point density correlation 
function $Q(t)$ goes to zero within the simulation time. For both the model systems, we have performed 
simulations for pin concentration in the range $\rho_{pin} \in \{0.005, 0.200\}$ for each temperature.
For very low temperatures, we were not able to equilibrate the system for high pin concentrations
because of a dramatic increase in the relaxation time in these cases.

\section{Typical overlap correlation functions}
\begin{figure}
\begin{center}
\includegraphics[height=.60\columnwidth]{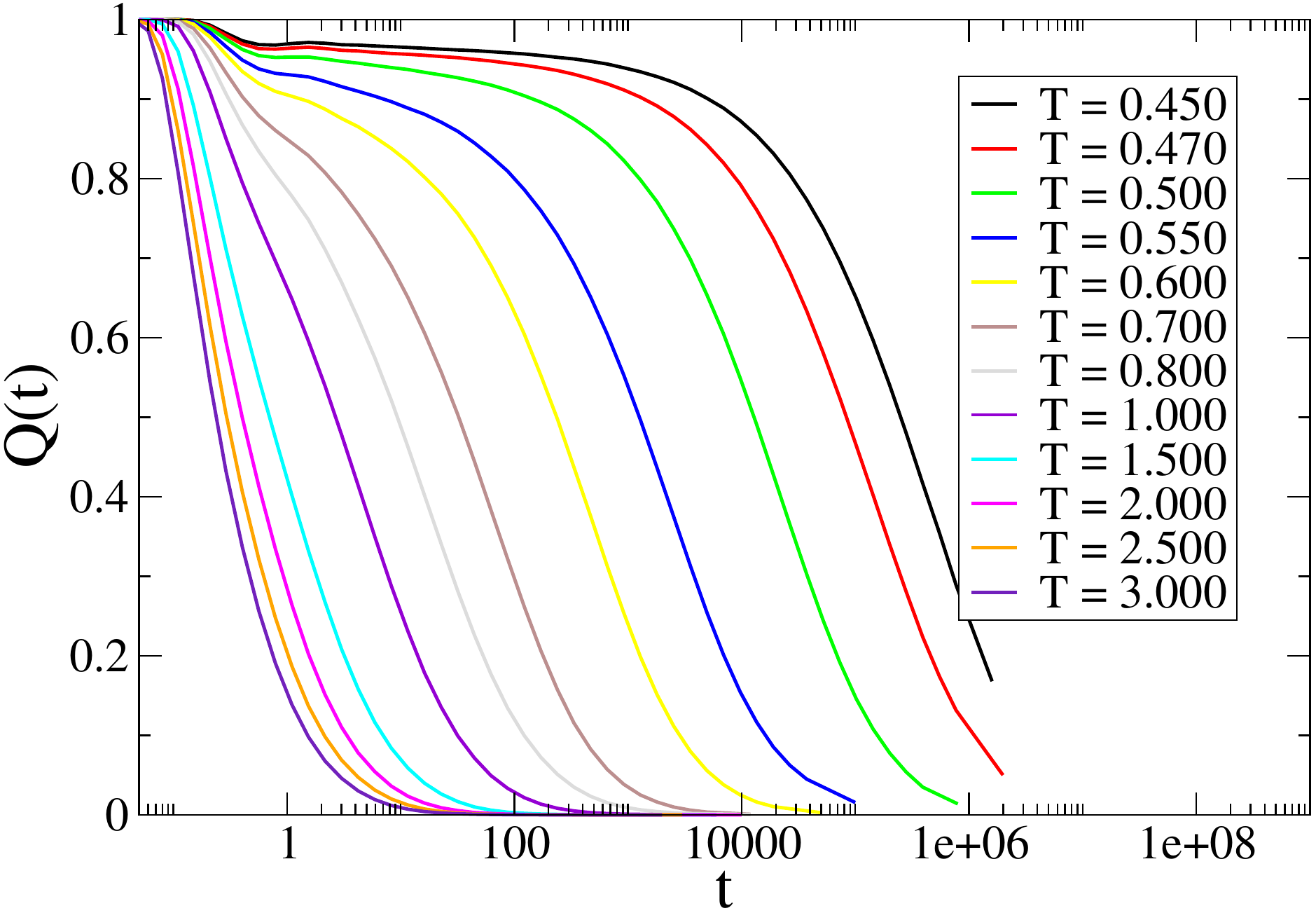}
\\
\vspace{.1\columnwidth}
\includegraphics[height=.60\columnwidth]{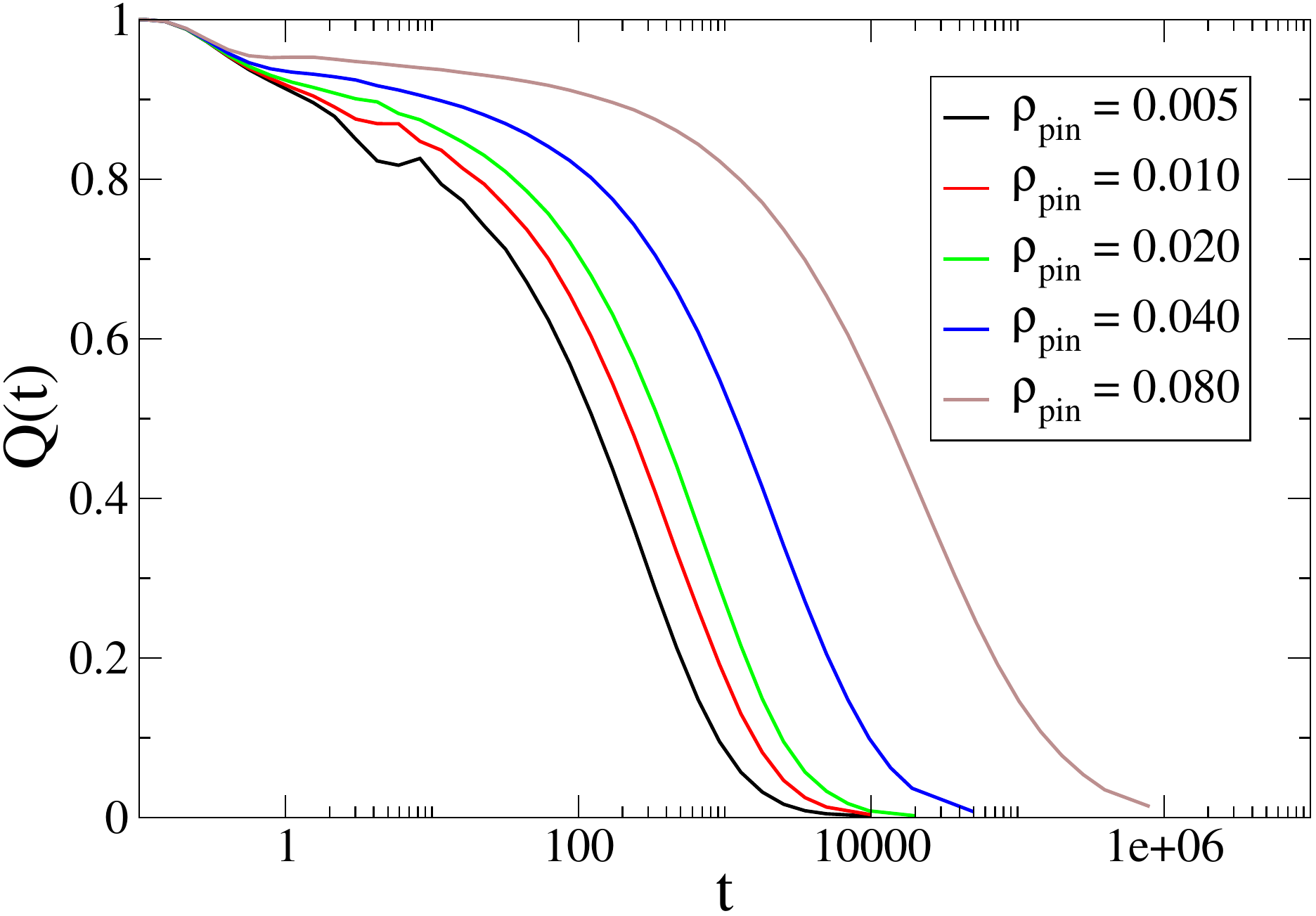}
\caption{{\em Top Panel:} $Q(t)$ for $\rho_{pin}=0.080$ for
  various temperatures for the $3dKA$ model.
  {\em Bottom Panel:} $Q(t)$ for $T=0.500$ for
  various levels of pinning for the $3dKA$ model.
\label{avgQts3dKA}}
\end{center}
\end{figure}

\begin{figure*}
\begin{center}
\vspace{1.2cm}
\includegraphics[height=.6\columnwidth]{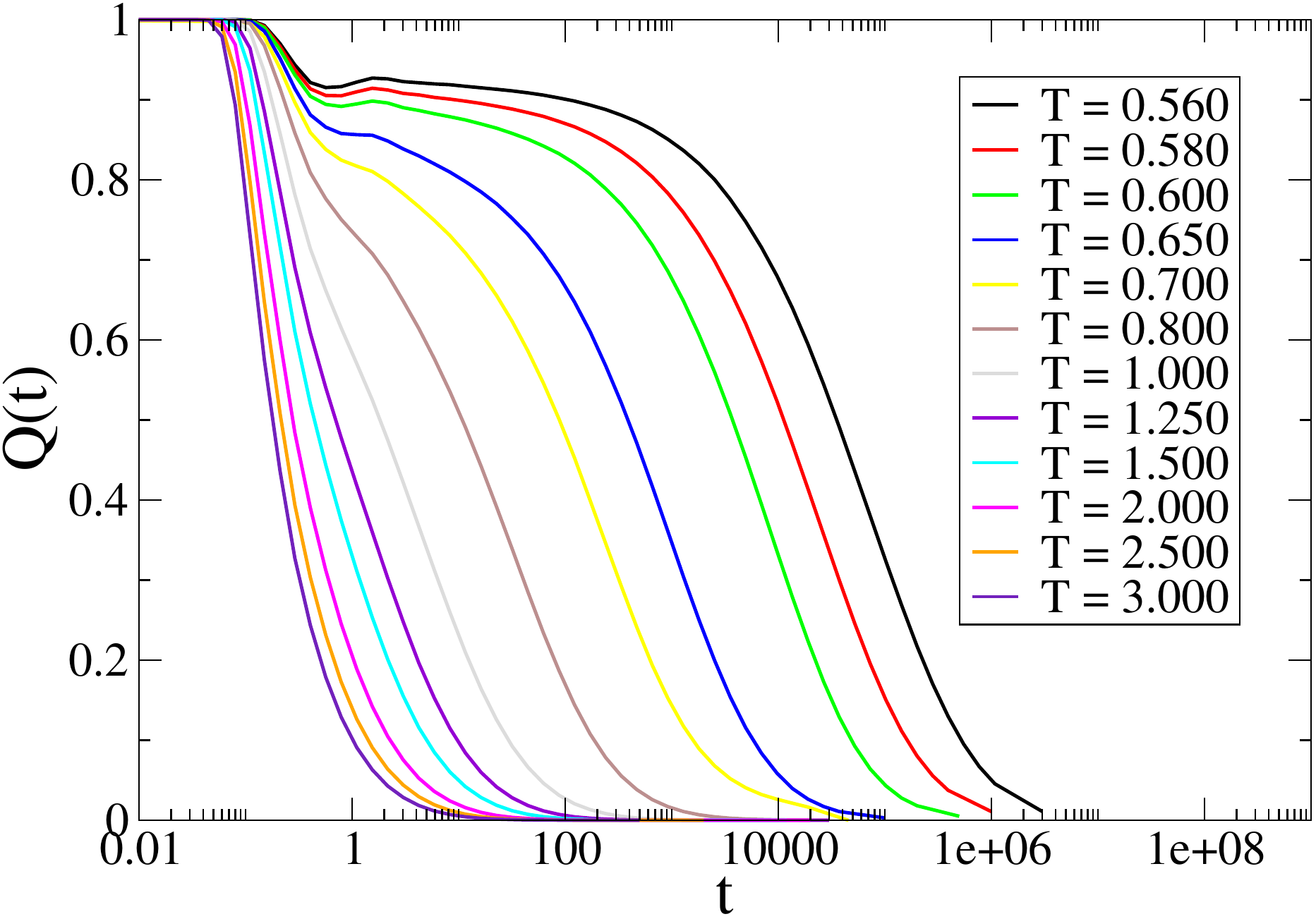}
\hspace{.1\columnwidth}
\includegraphics[height=.6\columnwidth]{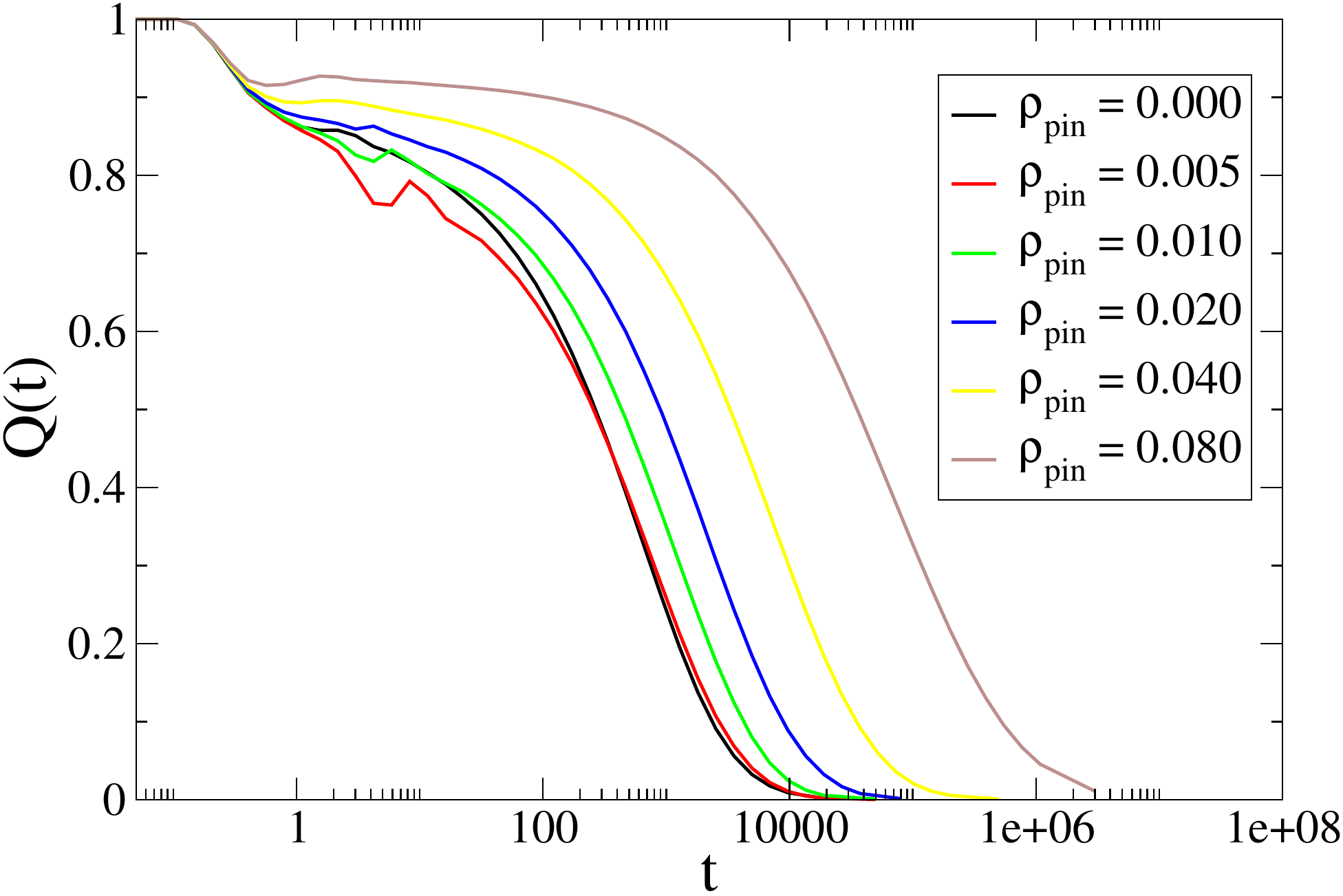}
\caption{{\em Left Panel:} $Q(t)$ for $\rho_{pin}=0.080$ for
  various temperatures for the $3dR10$ model.
  {\em Right Panel:} $Q(t)$ for $T=0.560$ for
  various levels of pinning for the $3dR10$ model.
\label{avgQts3dR10}}
\end{center}
\end{figure*}

Some typical overlap correlation functions $Q(t)$, from our
simulations have been plotted in Figs. \ref{avgQts3dKA} and 
\ref{avgQts3dR10}. The set has
data covering the entire range of relaxation times we obtained. The
relaxation time $\tau_\alpha$ was calculated using $Q(\tau_\alpha)=\exp(-1)$.

\begin{figure*}
\begin{center}
\vspace{0.2cm}
\includegraphics[height=.66\columnwidth]{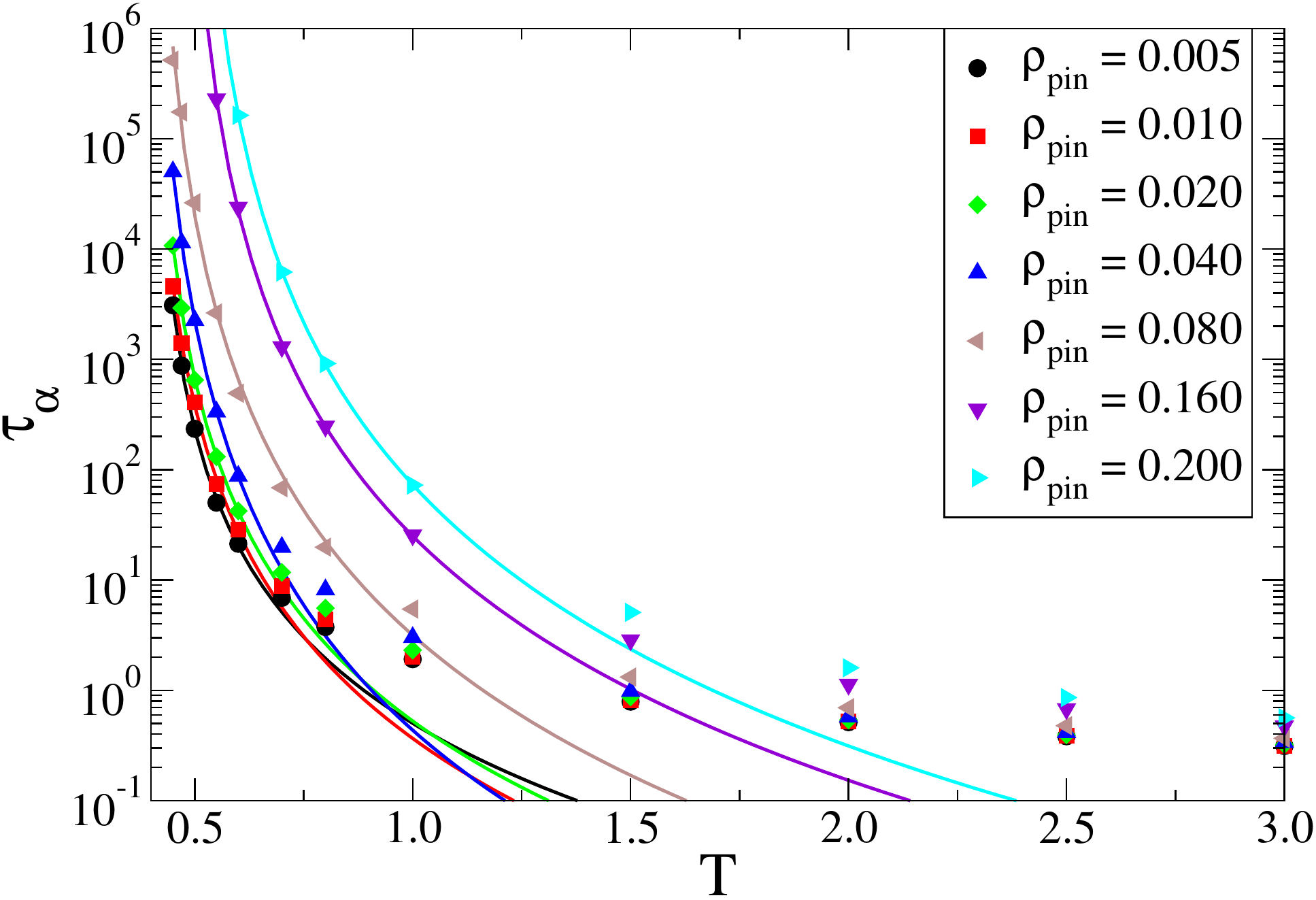}
\hspace{.1\columnwidth}
\includegraphics[height=.66\columnwidth]{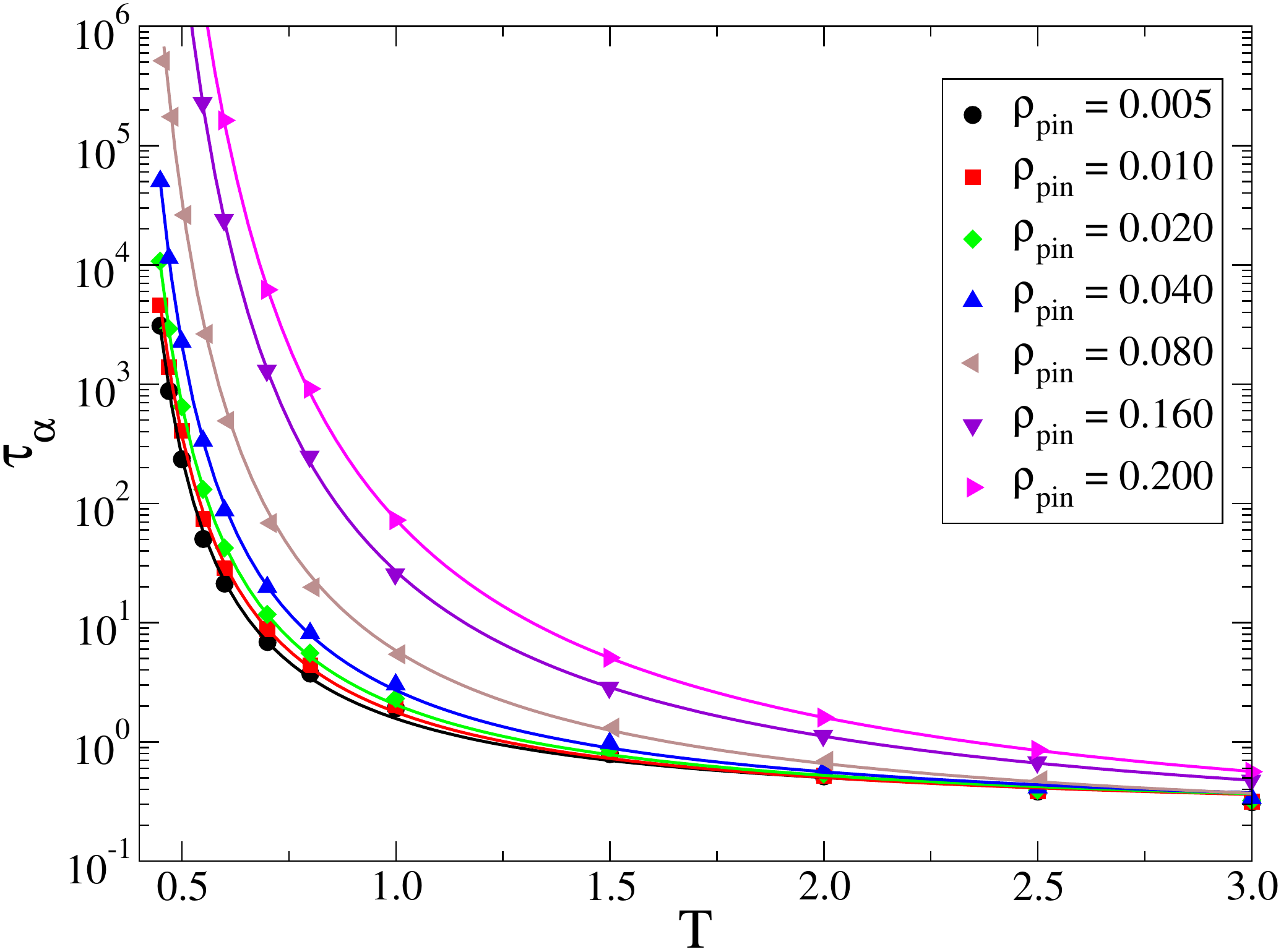}
\caption{{\em Left Panel:} Power law fit to obtain $T_C$ as a function
  of $\rho_{pin}$ for the $3dKA$ model.
  {\em Right Panel:} VFT fit to obtain $T_K$ as a function of
  $\rho_{pin}$ for the $3dKA$ model.
\label{mctVftFit3dKACurves}}
\end{center}
\end{figure*}

\begin{figure*}
\begin{center}
\vspace{0.3cm}
\hspace{0.0\columnwidth}
\includegraphics[height=.65\columnwidth]{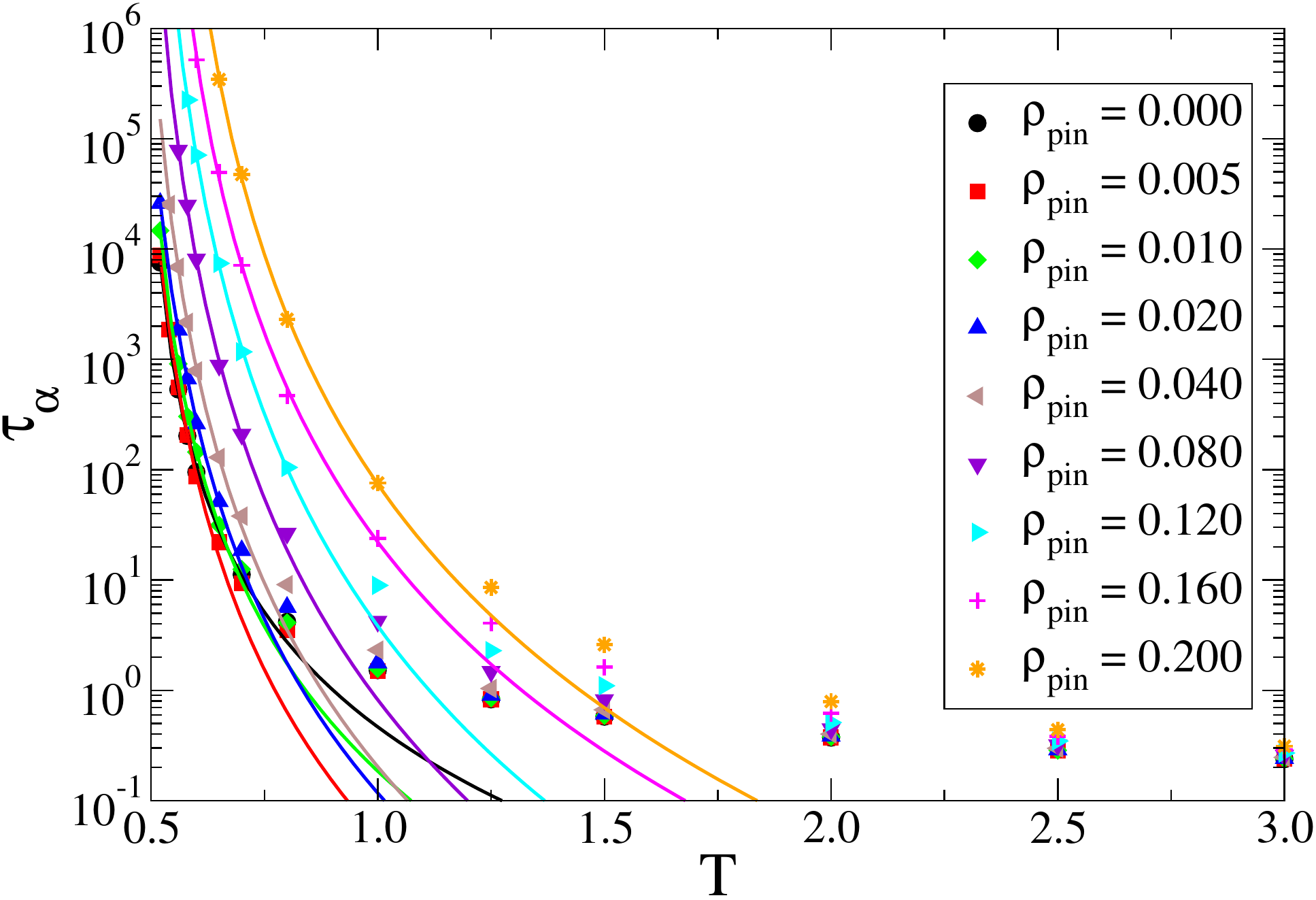}
\hspace{.1\columnwidth}
\includegraphics[height=.65\columnwidth]{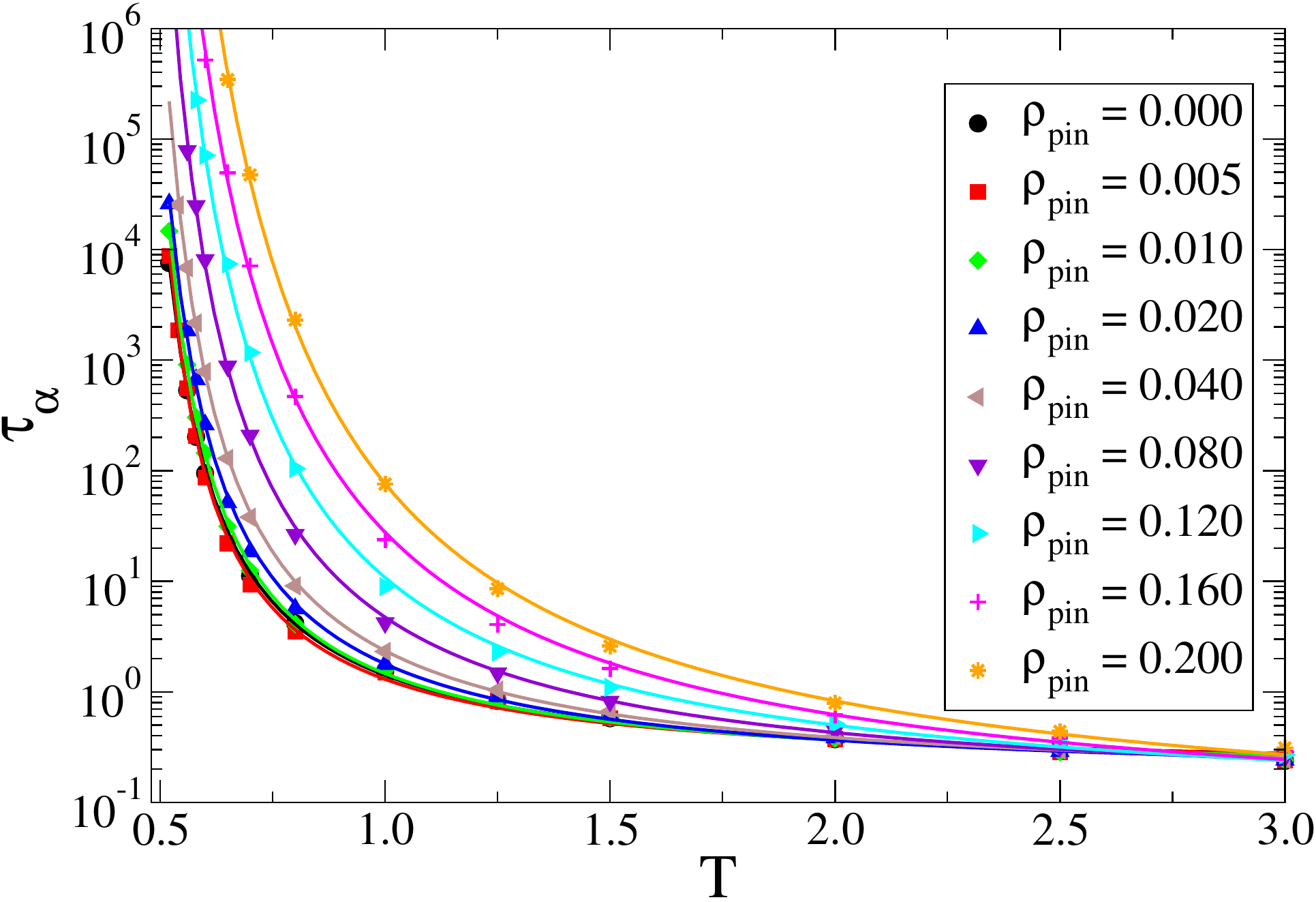}
\caption{{\em Left Panel:} Power law fit to obtain $T_C$ as a function
  of $\rho_{pin}$ for the $3dR10$ model.
  {\em Right Panel:} VFT fit to obtain $T_K$ as a function of
  $\rho_{pin}$ for the $3dR10$ model.
\label{mctVftFit3dR10Curves}}
\end{center}
\end{figure*}
\begin{figure}
\begin{center}
\vspace{0.2cm}
\includegraphics[height=.69\columnwidth]{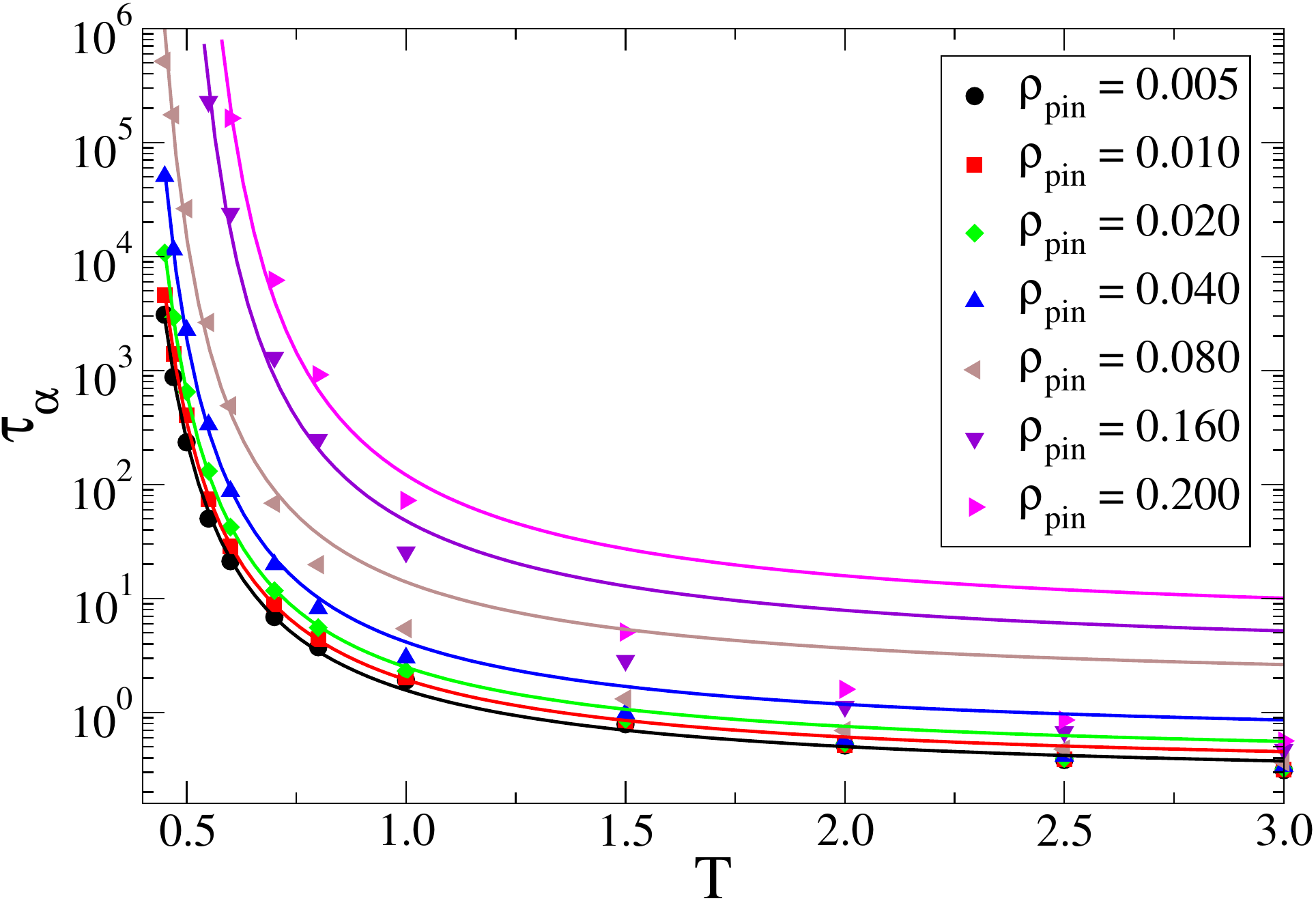}
\vskip 1.0cm
\includegraphics[height=.69\columnwidth]{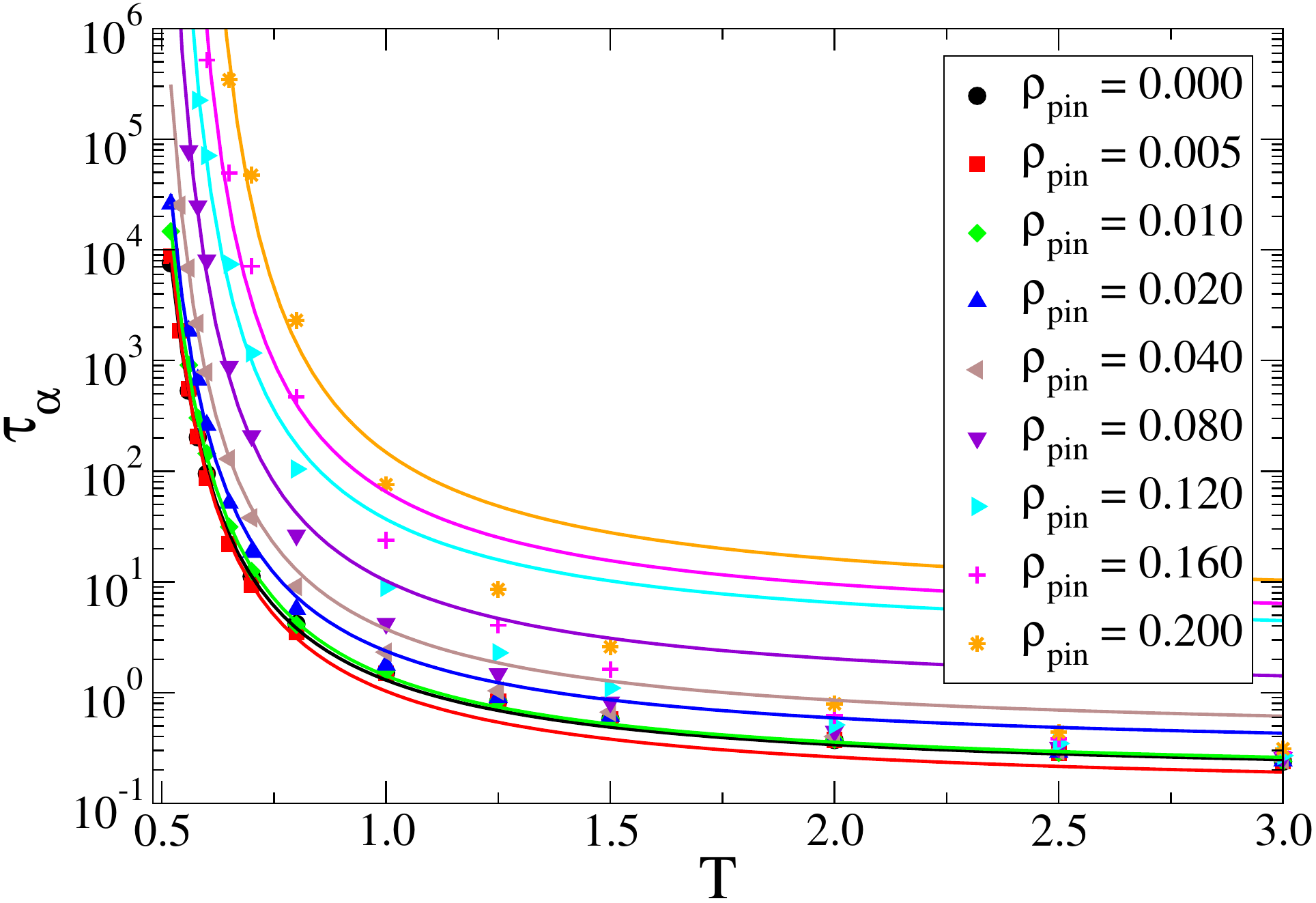}
\caption{{\em Left Panel:} VFT fit with fixed fragility for the $3dKA$ model.
  {\em Right Panel:} VFT fit with fixed fragility for the $3dR10$ model.
\label{fixedFragilityCurves}}
\end{center}
\end{figure}

\section{Fitting procedure}
The fitting was done using the method of least squares. The natural
logarithm of the relaxation time, $\ln(\tau_\alpha)$ was calculated first and
the fitting was done using the $\ln(\tau_\alpha)$ versus $T$ data.
The error bars for the parameters obtained from the fit were
calculated from the error in the $\tau_\alpha$-data using the
following procedure. For each value of $\tau_\alpha$, we generated
uniformly distributed data lying 
within the full 
width at half maximum of a Gaussian distribution with mean
$\tau_\alpha$ and variance $(\delta\tau_\alpha)^2$
($\delta\tau_\alpha$ being the error in $\tau_\alpha$).
Fits were done to 100 independent sets of such data. The spread in the parameters
obtained from these fits were taken as the errors in the fit parameters.

\begin{figure}
\begin{center}
\vspace{0.5cm}
\includegraphics[height=.69\columnwidth]{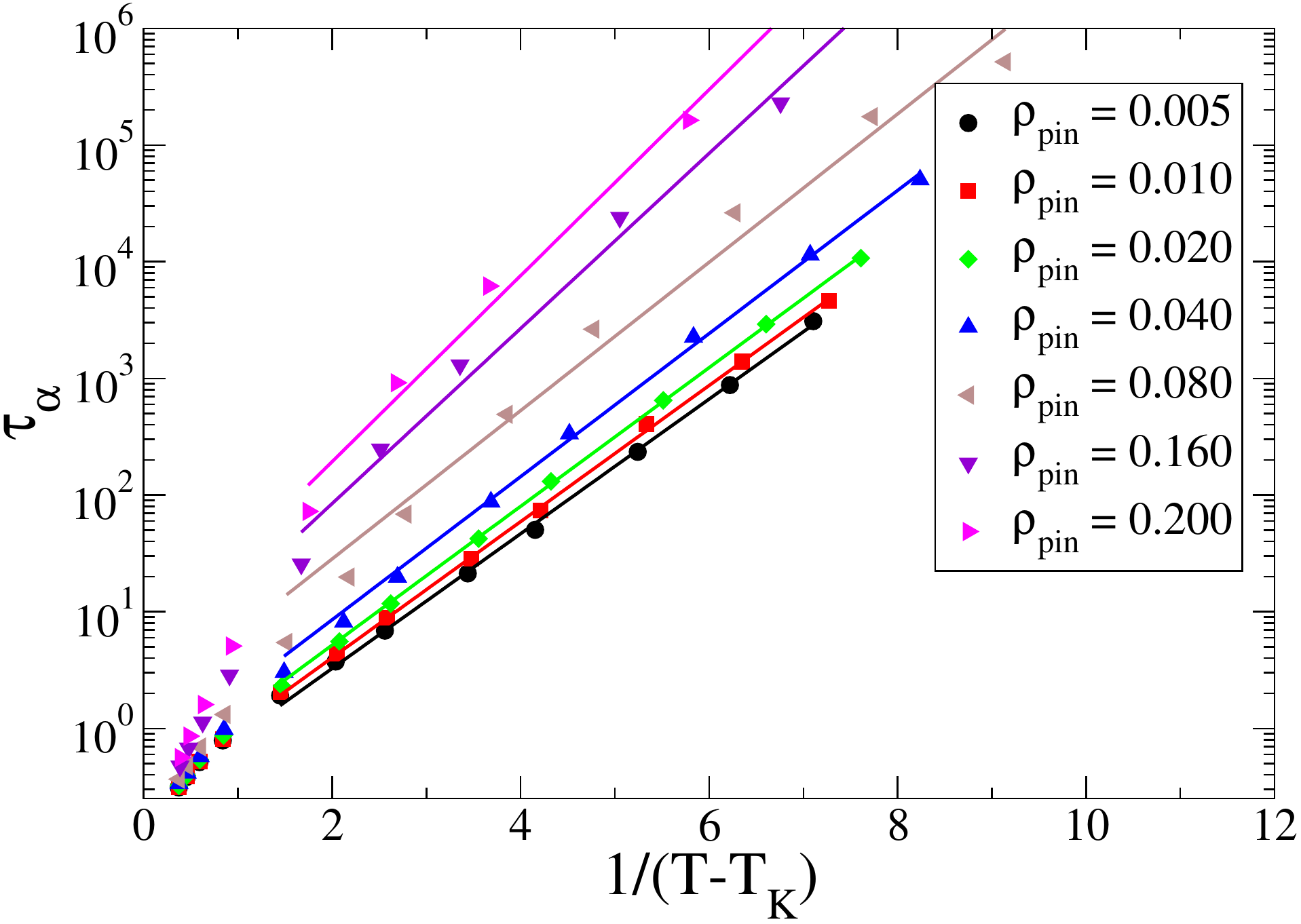}
\vskip +0.8cm
\includegraphics[height=.69\columnwidth]{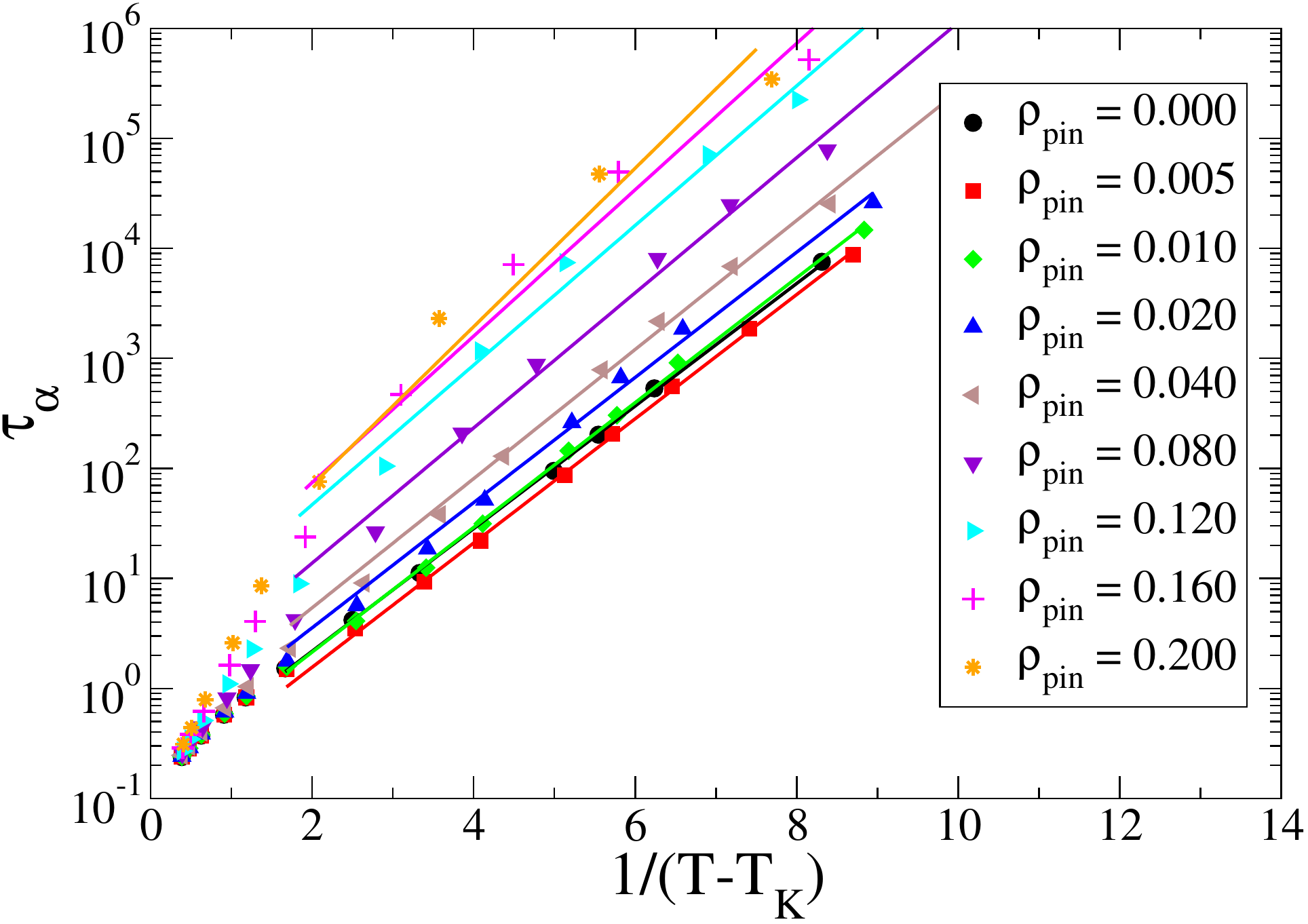}
\caption{{\em Left Panel:} Straight line plots for the VFT fit with
  fixed fragility for the $3dKA$ model. 
  {\em Right Panel:} Straight line plots for the VFT fit with fixed
  fragility for the $3dR10$ model. 
\label{fixedFragilityLines}}
\end{center}
\end{figure}
The power-law and VFT fits for the two studied models are shown in
Figs. \ref{mctVftFit3dKACurves} and \ref{mctVftFit3dR10Curves}. 
The same has been presented in the main paper in a way so that the
fits appear as straight lines.
The $T_K$ and $T_C$ values obtained from the fits are plotted in
Tables \ref{tktcTable3dR10} and \ref{tktcTable3dKA}.
\begin{table}
\begin{tabular}{| c | c | c | c |}
\hline
$\rho_{pin}$ & $T_K$ & $T_C$ \\
\hline
0.000&0.4012$\pm$0.0102&0.4852$\pm$0.0165\\
0.005&0.4122$\pm$0.0126&0.4567$\pm$0.0396\\
0.010&0.4052$\pm$0.0152&0.4721$\pm$0.0352\\
0.020&0.3942$\pm$0.0092&0.4492$\pm$0.0324\\
0.040&0.3992$\pm$0.0099&0.4595$\pm$0.0401\\
0.080&0.3872$\pm$0.0246&0.4643$\pm$0.0447\\
0.120&0.3722$\pm$0.0225&0.4740$\pm$0.0429\\
0.160&0.3502$\pm$0.0242&0.4996$\pm$0.0499\\
0.200&0.3432$\pm$0.0289&0.5171$\pm$0.0365\\
\hline
\end{tabular}
\caption{$T_K$ and $T_C$ values obtained from the VFT and MCT fits at
  various levels of pinning for the 3dR10 model.}
\label{tktcTable3dR10}
\end{table}

\begin{table}
\begin{tabular}{| c | c | c | c |}
\hline
$\rho_{pin}$ & $T_K$ & $T_C$ \\
\hline
0.005&0.3092$\pm$0.0098&0.4092$\pm$0.0163\\
0.010&0.2993$\pm$0.0092&0.3921$\pm$0.0310\\
0.020&0.2993$\pm$0.0112&0.4010$\pm$0.0300\\
0.040&0.2993$\pm$0.0095&0.3935$\pm$0.0157\\
0.080&0.2843$\pm$0.0173&0.4039$\pm$0.0208\\
0.160&0.2733$\pm$0.0349&0.4712$\pm$0.0377\\
0.200&0.2494$\pm$0.0528&0.4899$\pm$0.0270\\
\hline
\end{tabular}
\caption{$T_K$ and $T_C$ values obtained from the VFT and MCT fits at
  various levels of pinning for the 3dKA model.}
\label{tktcTable3dKA}
\end{table}

\subsection{Power law divergence}
The form assumed was
\begin{eqnarray}
\tau_\alpha=\frac{A}{(T-T_C)^\gamma},
\end{eqnarray}
where $A$ is a constant pre-factor having appropriate dimensions,
$\gamma$ is the exponent for the divergence, and $T_C$ is the (putative) 
transition temperature of mode coupling theory.
For each value of the pinning density $\rho_{pin}$, 4-5 data points
with the highest values of $\tau_\alpha$ were used for the fitting
procedure.

\subsection{VFT divergence}
The form assumed was
\begin{eqnarray}
\tau_\alpha=\tau_\infty\exp\left[\frac{1}{K_{VFT}\left(\frac{T}{T_K}-1\right)}\right],
\end{eqnarray}
where $\tau_\infty$ is the high temperature relaxation time, $K_{VFT}$
is the kinetic fragility and $T_K$ is the Kauzmann temperature.

\subsubsection{Fixed fragility}
\begin{figure}
\begin{center}
\includegraphics[height=0.66\columnwidth,angle=0]{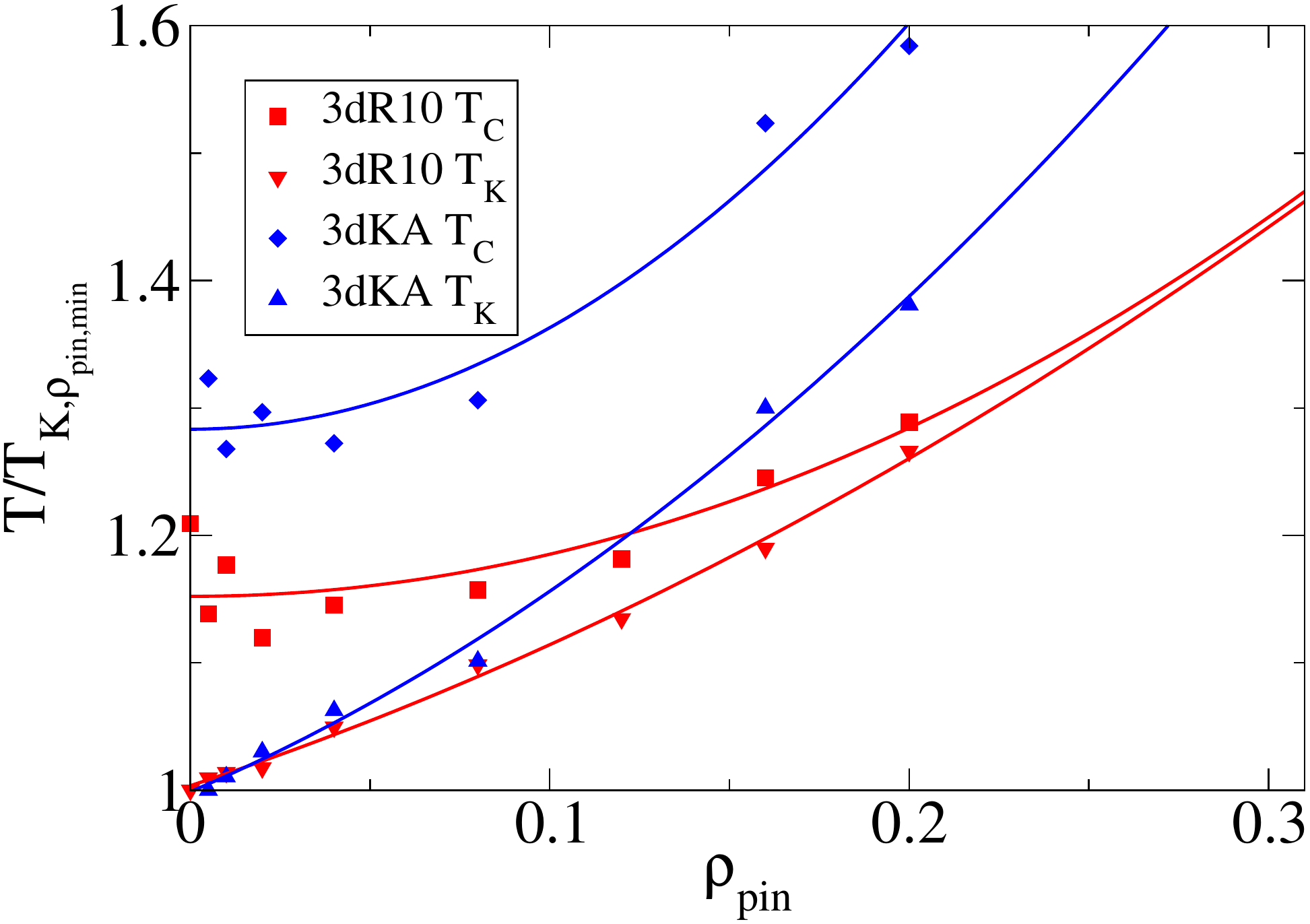}\hspace{.1\columnwidth}
\includegraphics[height=0.70\columnwidth]{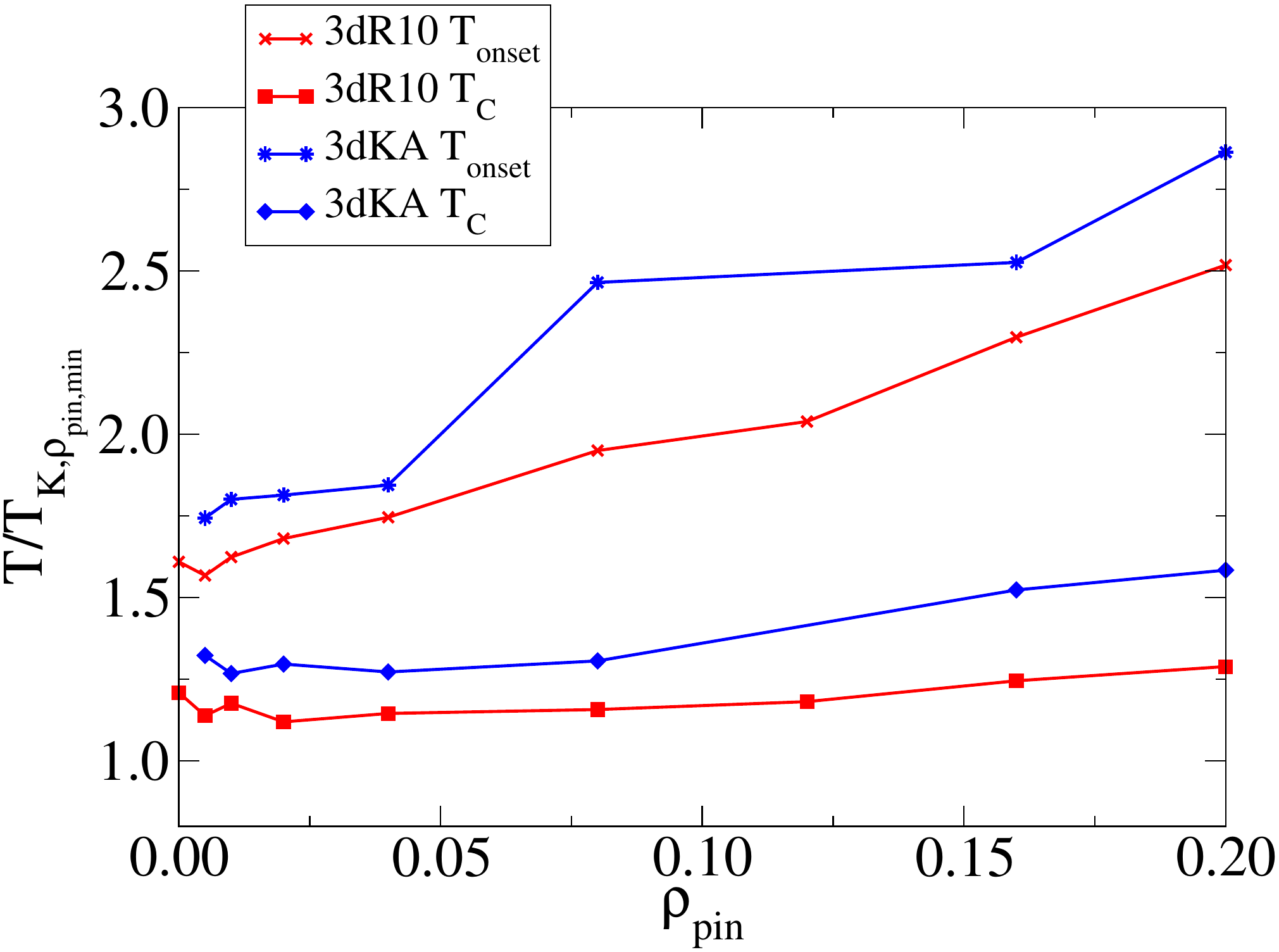}
\caption{{\em Left Panel:} The phase diagram with values of $T_K$
  obtained from fits with a fixed value of the 
  kinetic fragility. The lines are quadratic fits to the data. The
  fits for the $T_C$ data were done without a linear
  term.
  {\em Right Panel:} Variation of the 
  MCT-transition temperature $T_C$ and the onset temperature $T_{onset}$
  with $\rho_{pin}$ for the 3dKA and the 3dR10 models. 
\label{phaseDiagramsFxdFragOnset}}
\end{center}
\end{figure}

We obtained the same VFT fits also with fixed fragility. These are shown in
Figs. \ref{fixedFragilityCurves} and \ref{fixedFragilityLines}. Since
the fragility does change with pinning density, these fits were good
only in the low temperature regime. As such, for these fits, we only
used those data points for which $\tau_\alpha>10$.
The phase diagram obtained using these fits is shown in the left panel
of Fig. \ref{phaseDiagramsFxdFragOnset}.
\subsection{Onset temperature}
The onset of super-Arrhenius behavior is marked by the temperature
$T_{onset}$. This has been estimated by observing the meeting of the
Arrhenius fit to the high temperature data and the super-Arrhenius fit to
the low temperature data. This meeting point marks the crossover from
the high temperature Arrhenius behavior to the low temperature
super-Arrhenius behavior. The predictions based on MF and RFOT
analysis suggest that $T_{onset}$ should approach
$T_C$ as $\rho_{pin}$ is increased \cite{cammarotaPinning}. Our
results, plotted in the right panel of
Fig. \ref{phaseDiagramsFxdFragOnset}, do not support this 
prediction.

\begin{figure}
\begin{center}
\includegraphics[height=.50\columnwidth]{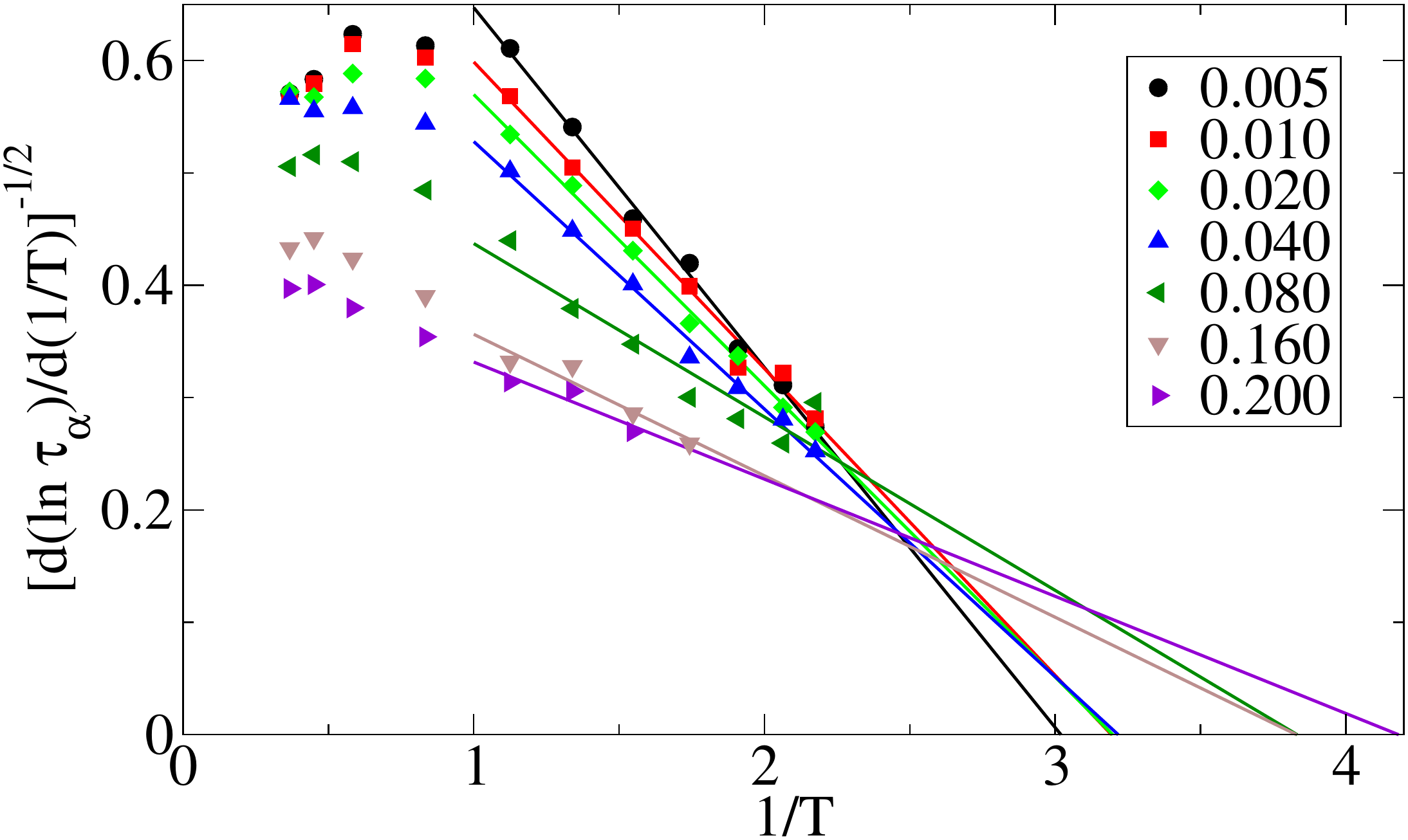}
\vskip +0.5cm
\includegraphics[height=.50\columnwidth]{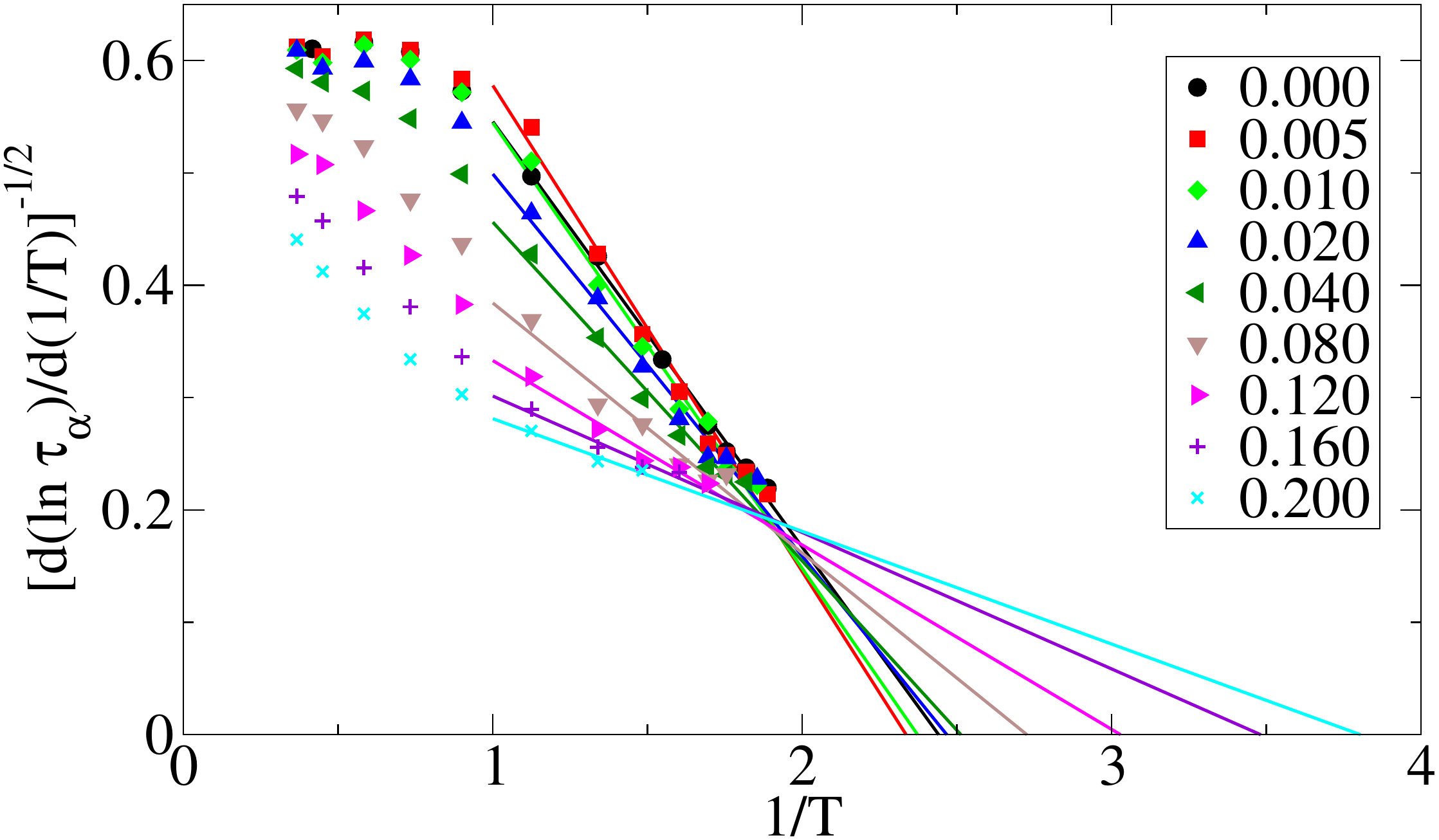}
\caption{Inverse temperature Stickel plots with linear fits in the
  region $T\le1$.  
  {\em Left Panel:} $3dKA$ model.
  {\em Right Panel:} $3dR10$ model.
  The $x$-intercepts give the values of $1/T_K$.
\label{stickelinvplotsvftfit}}
\end{center}
\end{figure}

\subsection{Stickel Plots}
An efficient way to verify the validity of the VFT form for the
divergence of the relaxation time at $T_K$ was  introduced in Ref. \cite{stickelplot}.
This method involves analyses of the derivative of the viscosity (or $\alpha$ relaxation time) with respect
to the temperature. A quantity ${\Phi}(T)$ is calculated using the data for the relaxation
times as follows.
\begin{eqnarray}
{\Phi}(T)
\equiv
\left(
\frac{d \ln \tau_\alpha}{d(1/T)}
\right)^{-\frac{1}{2}}.
\end{eqnarray}
It is easy to see
that 
\begin{eqnarray}
{\Phi}(T)
=
\sqrt{{K_{VFT}}{T_K}}\left(\frac{1}{T_K}-\frac{1}{T}\right),
\end{eqnarray}
if the VFT form holds. This implies that a plot of $\Phi(T)$ versus $1/T$ should be linear
and extrapolate to zero at $1/T_K$. Thus $T_K$ can be obtained from a linear fit to the $\Phi(T)$ versus $1/T$ 
data and the fragility parameter $K_{VFT}$ can be obtained from the slope of the fitted straight line.

Stickel plots (plots of $\Phi(T)$ versus $1/T$) for the two model systems considered here are shown
in Fig. \ref{stickelinvplotsvftfit}. It is clear from these plots, even without doing any fitting, that the data for 
larger values of $\rho_{pin}$ would extrapolate to zero at larger values of $1/T$, yielding lower values of $T_K$.
This is precisely the trend observed in the values of $T_K$ obtained from the VFT plots shown in the main paper.
Also, the Stickel plots in Fig. \ref{stickelinvplotsvftfit} do not show any indication of this trend 
changing at still lower temperatures.

\begin{figure}
\begin{center}
\vskip 0.5cm
\includegraphics[height=.45\columnwidth]{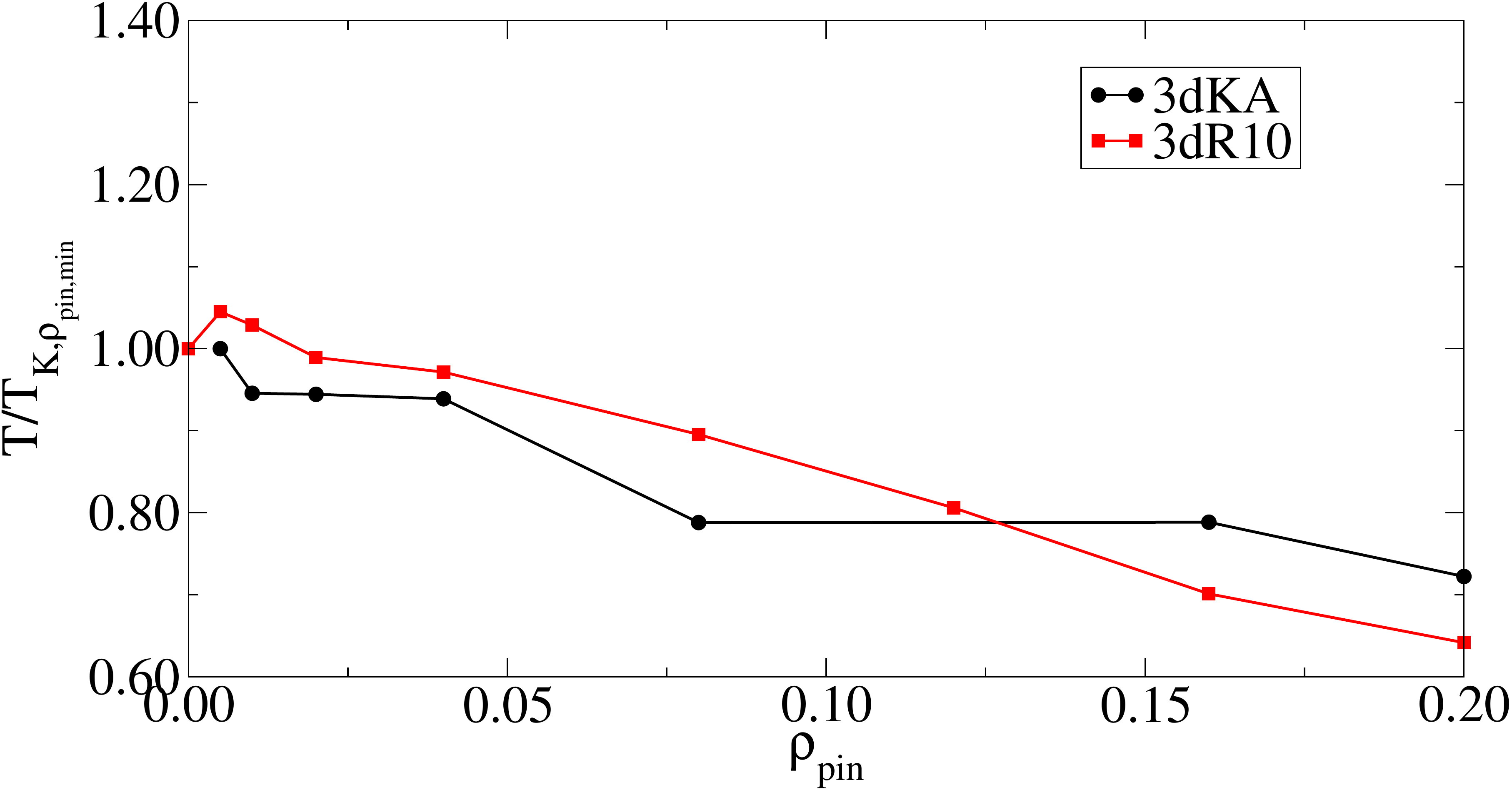}
\vskip + 1.0cm
\includegraphics[height=.6\columnwidth]{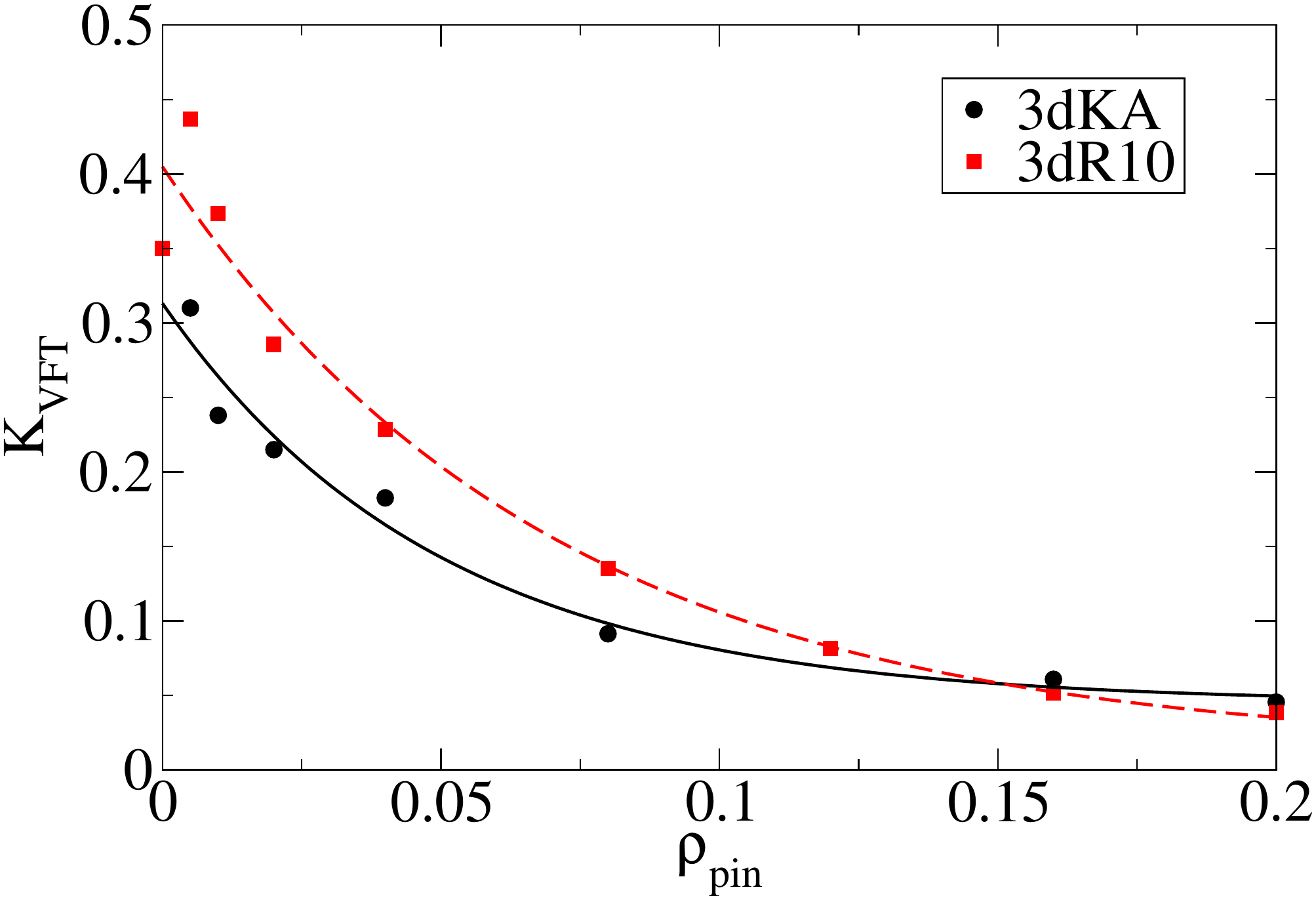}
\caption{Fit parameters obtained from fits in Fig. \ref
  {stickelinvplotsvftfit}.
  {\em Left Panel:} $T_K$.
  {\em Right Panel:} $K_{VFT}$.
\label{stickelinvplotsvftfitpars}}
\end{center}
\end{figure}

We fitted the low temperature parts ($T\le1$) of the ${\Phi}(T)$ data to
straight lines to get
$T_K$ and $K_{VFT}$. These fits are shown in
Fig. \ref{stickelinvplotsvftfit}. The values obtained 
(shown in Fig. \ref{stickelinvplotsvftfitpars}) agree with 
those obtained from our direct fits within error bars and show exactly the same general dependence on 
the pinning density.
